\documentclass[useAMS,usegraphicx,usenatbib]{mn2e} 
\usepackage{amsmath,graphicx}
\usepackage{subfigure}
\usepackage{ulem}
\usepackage{color}
\usepackage{graphicx}
\usepackage{times} 
\usepackage{amssymb}
\usepackage{amsmath}
\usepackage{lscape}
\usepackage{url}
\usepackage{multirow}
\usepackage{ulem}
\newif\ifAMStwofonts
\AMStwofontstrue
\def\epicmos1{{\it EPIC}{\rm-MOS1}}
\def\epicmos2{{\it EPIC}{\rm-MOS2}}
\def\epicmos{{\it EPIC}{\rm-MOS}}

\def\daophot{\hbox{\rm DAOPHOT}}

\def\daofind{\hbox{\rm DAOFIND}}

\def\cm{\hbox{$\rm\thinspace cm$}}

\def\kpc{\hbox{$\rm\thinspace kpc$}}

\def\pcmsq{\hbox{$\rm\thinspace cm^{-2}$}}
\def\pcmcu{\hbox{$\rm\thinspace cm^{-3}$}}
\def\pcmcuK{\hbox{$\rm\thinspace cm^{-3}~K$}}

\newcommand{\g}{\rm\thinspace g}

\def\gpcmsq{\hbox{$\g\cm^{-2}\,$}}

\def\n{\hbox{\rm NGC 1275}}

\begin{document}

\title[Star Formation in \n] {
Star Formation in the Outer Filaments of NGC 1275} \author[R.E.A.Canning \textit{et al.}] {\parbox[]{6.in}
  { R.~E.~A.~Canning$^{1}$\thanks{E-mail:
      bcanning@ast.cam.ac.uk}, A.~C.~Fabian$^{1}$, R.~M.~Johnstone$^{1}$, J.~S.~Sanders$^{1}$, C.~J.~Conselice$^{2}$, C.~S.~Crawford$^{1}$, J.~S.~Gallagher~III$^{3}$ and E.~Zweibel$^{3,4}$\\ } \\
  \footnotesize
  $^{1}$Institute of Astronomy, Madingley Road, Cambridge, CB3 0HA\\
  $^{2}$University of Nottingham, School of Physics \& Astronomy, Nottingham NG7 2RD\\
  $^{3}$Department of Astronomy, University of Wisconsin, Madison, Wisconsin 53706, USA\\
  $^{4}$Department of Physics, University of Wisconsin, Madison, Wisconsin 53706, USA}
\maketitle

\begin{abstract} 
We present photometry of the outer star clusters in NGC 1275, the brightest galaxy in the Perseus cluster. The observations were taken using the Hubble Space Telescope Advanced Camera for Surveys. We focus on two stellar regions in the south and south-east, far from the nucleus of the low velocity system ($\sim22\,\mathrm{kpc}$). These regions of extended star formation trace the H$\alpha$ filaments, drawn out by rising radio bubbles. In both regions bimodal distributions of colour ($B-R$)$_{0}$ against magnitude are apparent, suggesting two populations of star clusters with different ages; most of the H$\alpha$ filaments show no detectable star formation. The younger, bluer population is found to be concentrated along the filaments while the older population is dispersed evenly about the galaxy. We construct colour-magnitude diagrams and derive ages of at most $10^{8}\,\mathrm{years}$ for the younger population, a factor of 10 younger than the young population of star clusters in the inner regions of NGC 1275. We conclude that a formation mechanism or event different to that for the young inner population is needed to explain the outer star clusters and suggest that formation from the filaments, triggered by a buoyant radio bubble either rising above or below these filaments, is the most likely mechanism.
\end{abstract}


\section{Introduction}

\begin{figure*}
\centering
\includegraphics[width=\textwidth]{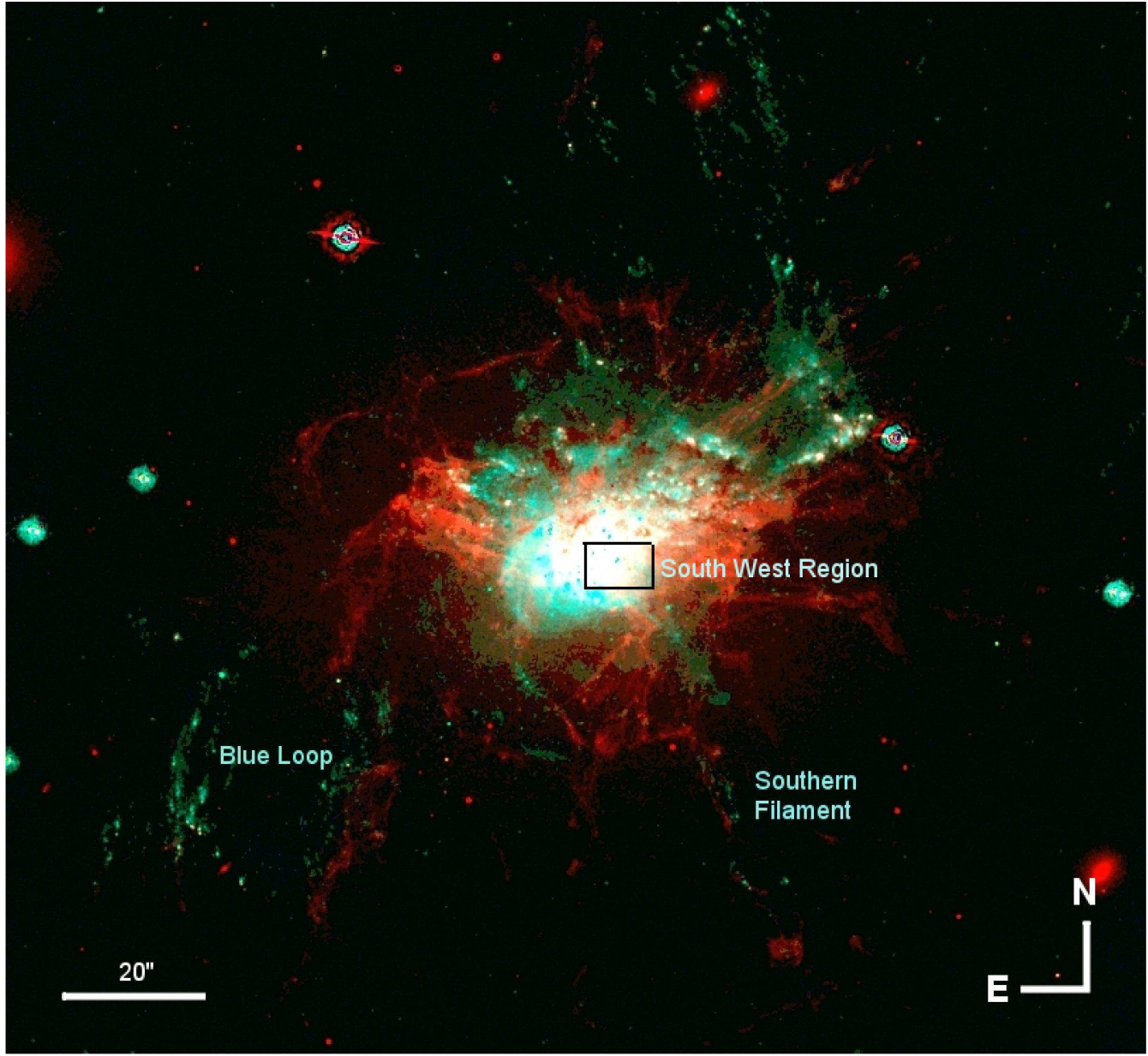}
\caption{Two colour image of NGC 1275. The green is a broad band B filter (F435W) and the red a broad band R filter (F625W) with the subtracted scaled green filter (F550M) image removing the contribution from the galactic continuum. The two regions of extended star formation discussed in this paper, the Blue Loop and the Southern
filament are labelled.}
\end{figure*}

Brightest Cluster Galaxies (BCGs), the most massive galaxies known, offer the 
opportunity to observe the heating and cooling processes in the Intra-Cluster
Medium (ICM) and provide the ideal environment to test theories of massive
galaxy formation. 

Optical line-emitting nebulae surround about a third of all BCGs
and are found specifically in those where the cluster exhibits a strongly peaked 
X-ray surface brightness profile and a cool, high density core, so called, `cool 
core' clusters (see \citealt{crawford2004} for a discussion of the optical properties
of these clusters).
The mass of cool gas implied by the high X-ray luminosity (i.e. the energy loss rate of the gas) in the centre of cool core clusters is ten or more times the mass inferred from soft X-rays ($<1\,\mathrm{keV}$), observed star formation rates and the mass of cold gas present \citep{peterson2006}.
In order for the hot gas to be radiating but not cooling in the predicted quantities 
there must be a source of heat regulating the production of cool gas.

Energy arguments favour feedback from a SuperMassive Black Hole (SMBH), situated in the central galaxy, as the dominant heating process in these clusters (for a review see \citealt{peterson2006, mcnamara2007}). Understanding the role the black hole plays with respect to the galaxy's environment such as
its connection with star formation, is a key question in galaxy evolution.

NGC 1275 is the BCG in the nearby ($z=0.0176$) Perseus Cluster, the X-ray brightest, cool core cluster. The NGC 1275 system is very complex;
\cite{minkowski1955} discovered it
consists of two galaxies, a high velocity system (HVS, $8200\,\mathrm{km}\,\mathrm{s}^{-1}$) and a low velocity system (LVS, $5200\,\mathrm{km}\,\mathrm{s}^{-1}$). 21 cm observations, line absorption in HI and Ly$\alpha$ and continuum absorption in X-rays \citep{ekers1976, briggs1982, fabian2000, gillmon2004} show the HVS to lie in front of the LVS. The HVS can be clearly seen in Fig. 1, north-west of the centre.

Early observations showed a spatial correspondence between the HVS and LVS \citep{hu1983, unger1990} and evidence for gas at intermediate velocities \citep{ferruit1997}, which led to the suggestion that the galaxies were in the process of merging. However, these observations can also be explained by influences of the ICM \citep{fabian1977, boroson1990, caulet1992} or by a past interaction, of one of the systems, with a third gas-rich galaxy \citep{holtzman1992, conselice2001}. By examining the X-ray absorption, \cite{gillmon2004} were able to put a lower limit on the distance between the HVS and the nucleus of the LVS, of $57\,\mathrm{kpc}$. This analysis was repeated with a deeper study by \cite{sanders2007} giving an improved lower limit on the distance between the two galaxies as $110\,\mathrm{kpc}$. This large separation and lack of obvious shocked gas suggests that any interaction between these two systems lies in their future not in their past.

NGC 1275 exhibits interesting optical features in both its relatively blue 
central colours \citep{holtzman1992} indicating the presence of massive, short-lived stars and in its vast 
extended emission-line nebulae \citep{minkowski1957, lynds1970} stretching out predominantly radially from the nucleus of the LVS. Large quantities of molecular hydrogen and CO gas has been found in the nuclear region \citep{lazareff1989, inoue1996, donahue2000} and recently detected in the outer H$\alpha$ filaments \citep{hatch2005, salome2008a, lim2008, ho2009}. Soft X-ray emission has also been found to be associated with some of the optical emission nebulae \citep{fabian2003}, however is less luminous than the optical and UV emission.

\cite{kent1979} found the low resolution spectra of the H$\alpha$ filaments to be similar to those of H II regions and well explained by collisional ionisation or photoionisation from the Seyfert nucleus. Their observations supported a hypothesis that the filaments were formed by an accretion flow onto the LVS. It has since been shown, with higher resolution spectra, that the galaxy nucleus is not the ionisation source \citep{johnstone1988}.
\cite{heckman1989} observed both the filaments in NGC 1275 and broadened their study to include filamentary systems in other BCGs. They found evidence confirming the association of optical emission-line nebulae and cooling flows and discuss mechanisms for heating and ionising these nebulae.

The central star cluster population in NGC 1275 has been well studied;
\cite{shields1990} detected H II regions and excess blue continuum in the central filaments of NGC 1275 which they interpreted as implying the presence of massive young star clusters. Soon after, \cite{holtzman1992} discovered a population of 
compact massive blue star clusters in the core of NGC 1275 with the Hubble
Space Telescope (HST) Wide 
Field Planetary Camera 1 (WF/PC1). The observed sizes and luminosities led the authors to conclude 
that these were 
most likely proto-globular clusters. The conclusion was supported by \cite{richer1993} using observations from the Canada-France-Hawaii Telescope (CFHT) and by \cite{carlson1998}
using HST WFPC2 observations. Assuming a Salpeter 
Initial Mass Function (IMF), previous ages of the central clusters are suggested to be between $10^{8}-10^{9}$ years 
resulting in masses between $2\times10^{7}-10^{8}\,\mathrm{M}_{\odot}$. 

A further study of the colour distribution of the NGC 1275 system was carried out by \cite{mcnamara1996}. 
The $U-I$ colours indicate a centrally concentrated young population and a more diffuse older background population. 
The young population colours are consistent with ages up to $1\,\mathrm{Gyr}$. Measuring the physical parameters 
of star clusters in cD galaxies can allow
us to constrain both their rate of formation and the underlying physical processes
behind their development.

The complicated nature of the NGC 1275 system makes it difficult to determine whether the star cluster populations in the center lie 
in the LVS or HVS. \cite{keel2001} suggested that the correlated spatial distribution of the blue clusters and the HVS dust lanes are 
evidence of this cluster population being part of the foreground HVS. However, \cite{brodie1998} have shown that at least some of the 
blue clusters are at the same redshift ($\sim5200\,\mathrm{kms}^{-1}$) as the LVS. The colours of the blue population have been 
determined to be similar, regardless of whether the individual clusters have been confirmed to lie in the LVS or are perhaps, spatially, 
more likely to lie in the HVS \citep{holtzman1992, norgaardnielsen1993, richer1993, carlson1998}. \cite{dixon1996} observed NGC 1275 with 
the Hopkins Ultraviolet Telescope and found evidence for a star formation rate of 30$\mathrm{M}_{\odot}\,\mathrm{yr}^{-1}$ at the redshift of the HVS.

There are two major competing scenarios for the formation of Globular Clusters (GC) in
elliptical galaxies.
\cite{ashman1992} propose that elliptical galaxies form through
mergers which trigger the GC formation while \cite{forbes1997} proposes that the GCs form `in situ' via collapse into a single potential well (for a review see \citealt{brodie2006}).
The merger
scenario produces a bi-modal distribution of clusters and is supported by the observation 
that elliptical galaxies tend to have a higher specific frequency (number of clusters per unit galaxy light) than spiral
galaxies. The formation processes of the clusters in the centre of NGC 1275
are discussed by \cite{carlson1998} and \cite{richer1993}. The former
prefer a model where the star formation is triggered by a merger while the latter suggest the star formation is due to cool gas collected at the centre of the cooling flow. The
main contention is whether the colour-magnitude diagram shows evidence
of a single age population or of a continuously forming population of star clusters.

\cite{ferruit1994} used TIGER, an Integral Field Spectrograph (IFS) mounted on the $3.6\,\mathrm{m}$ CFHT to detect emission line regions coincident with the central clusters, previously assumed only continuum sources. These regions exhibit spectral features very different to the H II regions discovered previously and have spectral and kinematical properties characteristic of the filaments. The line ratios suggest that photoionisation by the star clusters cannot account for the gas ionisation in the filaments. These findings, coupled with the uncertainties of our knowledge of the internal reddening in NGC 1275, led the authors to support the theory that the clusters were formed from a cooling flow.

Further IFS observations with GMOS on GEMINI were taken by \cite{trancho2006}. These observations confirmed the discovery of emission lines associated with some star clusters, although the majority exhibit very little gas emission.

The star formation is however not limited to the central region (see regions marked on Fig. 1). NGC 1275
also exhibits regions of very extended star formation noted by \cite{sandage1971}. The `blue knots' of \cite{sandage1971} form the eastern side of the Blue Loop, approximately $22\,\mathrm{kpc}$ to the south-east of the nucleus (Fig. 1). \cite{conselice2001} presented colours and magnitudes for clusters
near the north-west of the HVS and a few bright clusters on the eastern arm of the Blue Loop. On the Blue Loop they measure very
blue colours with ($B-R$)$_{0}$ $\approx$ $-0.3$ to 0.0. 
\begin{figure}
\centering
\includegraphics[width=0.45\textwidth]{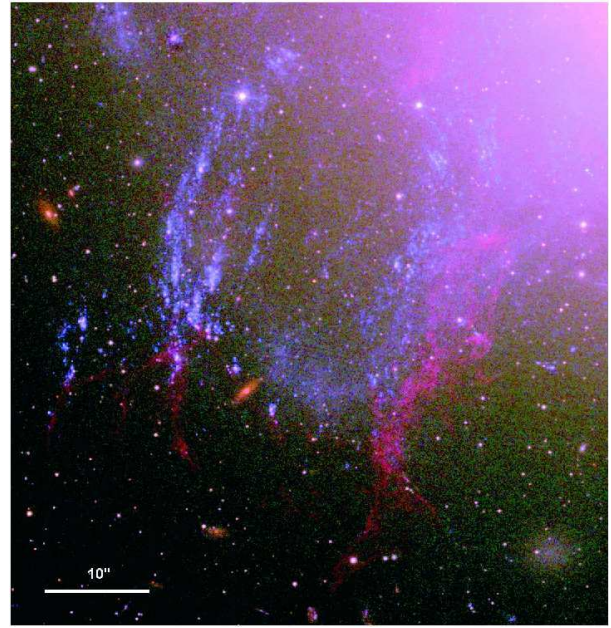}
\caption{Image showing the star clusters (blue) and the H$\alpha$ emission (red) that make up the Blue Loop. Image from \protect \cite{fabian2008}.}
\end{figure}

These objects can be
dated from their colour-magnitude relationships and through this we can examine their
relationship to the H$\alpha$ filaments and to the wider cooling and heating
flows in the galaxy.
Observations of cooling flows and star formation in BCGs can give us an
insight into the evolutionary cycle that the gas undergoes in these objects.   
\begin{figure*}
\centering
\includegraphics[width=0.7\textwidth]{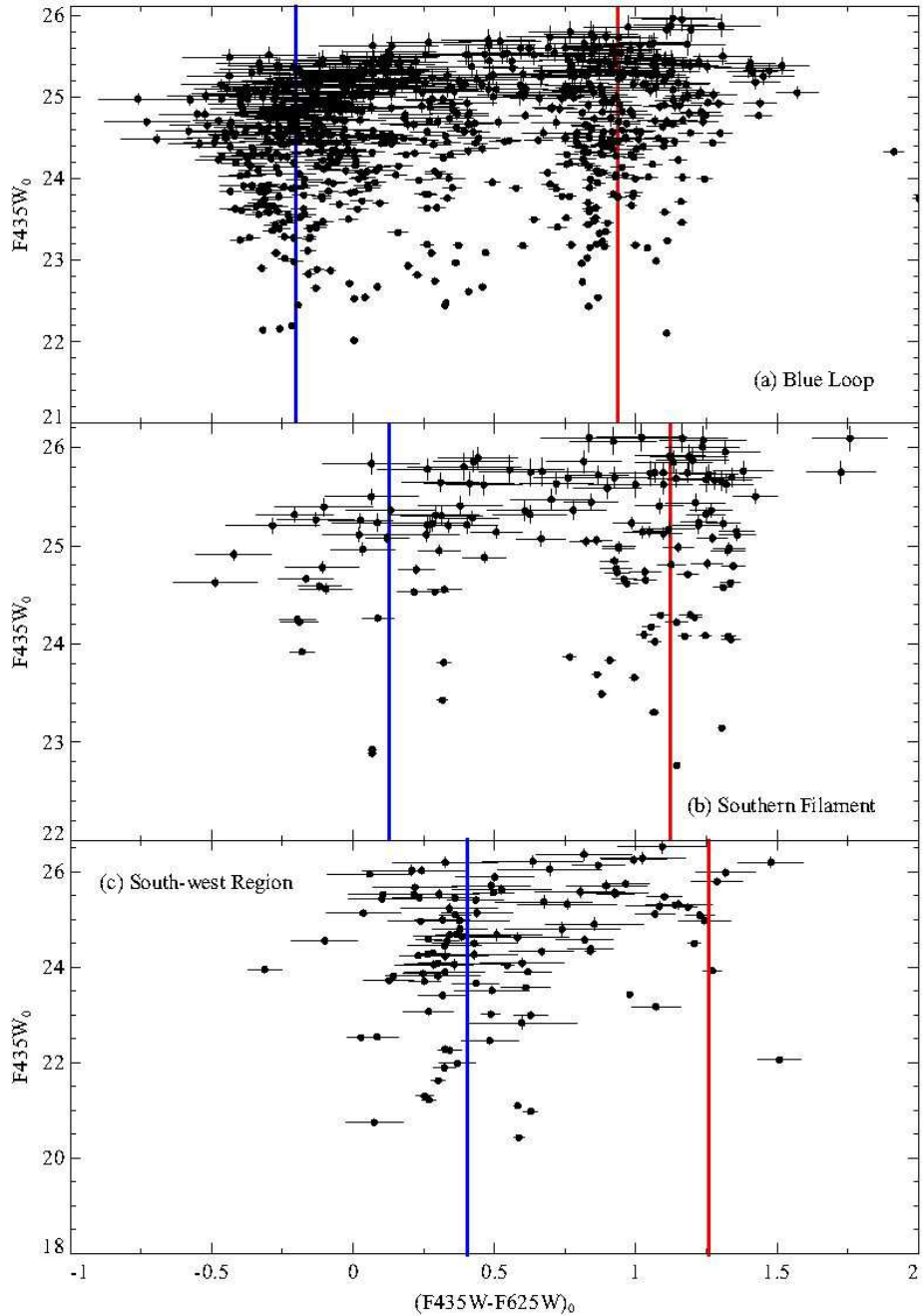}
\caption{Vega magnitude (F435W$_{0}$) \protect \footnotemark against colour (F435W-F625W)$_{0}$ for the three regions. All data points have 
errors in single passbands less than 0.15 mag. Panel (a) (Top) shows the 
Blue Loop region (RA $03^{h}19^{m}52^{s}.1$, Dec. $41^{\circ}$ 30'08.6'' J2000), panel (b) the Southern filament (RA $03^{h}19^{m}46^{s}.5$, Dec. $41^{\circ}$ 30'04.5'' J2000) and panel (c) the south-west section (RA $03^{h}19^{m}47^{s}.9$, Dec. $41^{\circ}$ 30'38.8'' J2000) of the nucleus. The blue and red lines in panels (a) and (b) indicate the mean colours of the blue and red populations respectively. The blue and red populations are determined by inspection of the histogram (see Fig. 4). The lines in panel (c) show the average colours determined by \protect \cite{carlson1998} for their blue and red populations, in the same region.}
\end{figure*}

In this paper we present observations, obtained with the HST Advanced Camera for Surveys (ACS) spanning an
area of approximately 5.5 square arcmin. The observations allow us to analyse the star formation spatially coincident with the H$\alpha$ emission line filaments in the outer regions of the system. The regions are too far from the centre of NGC 1275 to be seen on the WFPC2 images of \cite{carlson1998}. In Section 2 we discuss the observations and data
reduction procedure and in Section 3 present the results of the analysis and
highlight potential sources of error. We discuss the results in the context of
different formation scenarios of the extended star formation in Section 4 and conclude in Section 5.

\begin{figure*}
\centering
\includegraphics[width=0.8\textwidth]{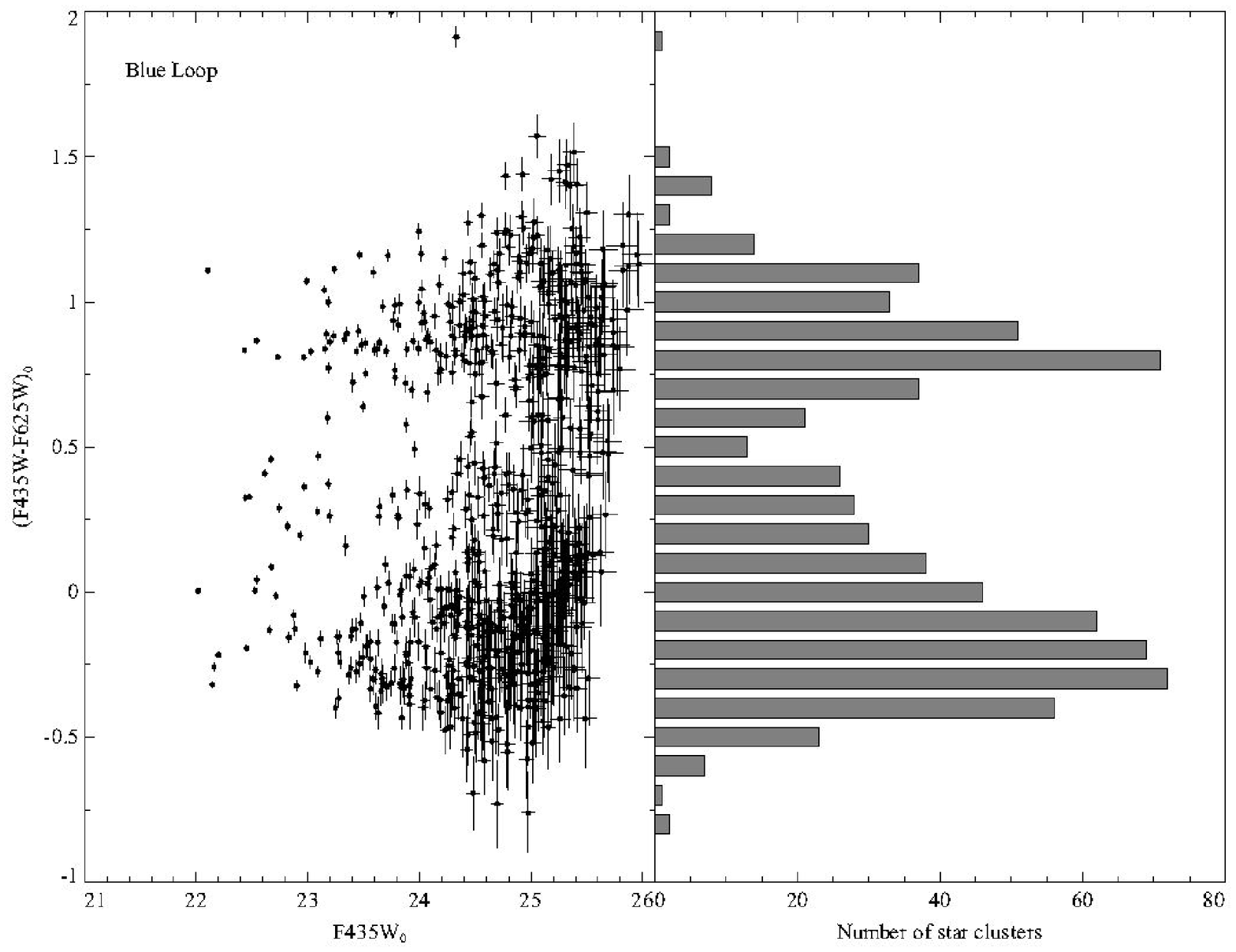}
\caption{(Left) The extinction-corrected colour-magnitude diagram for the Blue Loop. (Right) The number of clusters, in this region, detected in colour bins of width 0.1 mag. Here the bimodal distribution of colours can be clearly seen. The blue population (lower) has a broader distribution than the red population (upper), it also appears to have a tail towards the red end. The cut off between the two populations lies at $(\mathrm{F435W}-\mathrm{F625W})_{0} \sim 0.4$. The mean colour of the young, blue population is $(\mathrm{F435W}-\mathrm{F625W})_{0} = -0.22$.}
\end{figure*}
\section{Observations and Data Reduction}

\footnotetext{The $_{0}$ notation used for magnitudes and colours in this paper refer to corrections for foreground Galactic extinction only.}

Observations were made using the HST ACS on 2006 August 5th in the F435W, F550M and
F625W filters \citep{fabian2008}. Total exposure times were 9834, 12132 and 12405 s respectively. 

The F625W filter includes H$\alpha$, [N II] and [S II] emission from both the HVS and the LVS. This differs from the \cite{conselice2001} ground-based, narrow-band imaging that excluded the HVS.
\begin{table}
\begin{center}
\begin{tabular}{|l|l|l|l|}
\hline
Filter & \multicolumn{3}{c}{Aperture Correction} \\
\hline \hline
 & A &  B &  C \\
F435W & 1.07 & 1.06 & 1.05 \\
F550M & 1.07 & 1.06 & 1.04 \\
F625W & 1.08 & 1.07 & 1.04 \\
\hline
\end{tabular}
\end{center}
\caption{Aperture corrections for the three ACS filters used. A corresponds to the Blue Loop,
B the Southern filament and C the south-west portion of the nucleus. This correction was determined
from the 10 brightest sources in each region.}
\end{table}
\begin{table}
\begin{center}
\begin{tabular}{|l|l|}
\hline
Filter & Extinction Coefficient \\
\hline \hline
F435W & $A_{\mathrm{F435W}}=0.67$ \\
F550M & $A_{\mathrm{F550M}}=0.50$ \\
F625W & $A_{\mathrm{F625W}}=0.43$ \\
\hline
\end{tabular}
\end{center}
\caption{Galactic extinction coefficients for each ACS pass-band used.}
\end{table}
The standard procedure was used to bias-subtract and flat-field the data frames
which were then drizzled together as described in the HST ACS data handbook.

Fig. 1 shows the ACS F435W filter image of NGC 1275 combined with a red image made by subtracting the scaled F550M image from the broad band F625W filter, removing the contribution from the galactic continuum. Here the relation between the star formation and the line emission nebulae can be seen clearly. The HVS is in the line of sight of the nucleus of the LVS and appears to have a backward S shaped configuration.
Fig. 2 shows more clearly the relationship between 
the H$\alpha$ filament and the star clusters in the Blue Loop region. Here it can be seen 
that on the western arm the clusters are spatially offset from the filament by approximately 10 arcsec 
corresponding
to a distance of 3.5\kpc\ at a redshift of $z=0.0176$ (we adopt $H_{0}=71\,\mathrm{km}\,\mathrm{s}^{-1}\,\mathrm{Mpc}^{-1}$). Clusters on the eastern arm coincide directly (in the line of sight) with the H$\alpha$ filament.

We detected star clusters in the south-west region of NGC 1275 using IRAF \daofind\ and the method of
\cite{carlson1998}. We find that the clusters in the Blue Loop and Southern filament region are more 
compact than the clusters in the centre. The point-spread-function fitting technique
for crowded fields employed by IRAF \daophot\ was not used as many of the clusters are
partially resolved, and we required that a single technique should be used for 
the whole field.
As in \cite{carlson1998}, we
required the \daofind\ parameters of roundness between $-1$ and 1 and sharpness between 0.2 and 1. In the central image, only
the south-west region was considered; it is both less obviously `dusty' and
allows a more direct comparison of our colour-magnitude results with previous results.

\begin{figure}
\centering
\includegraphics[width=0.5\textwidth]{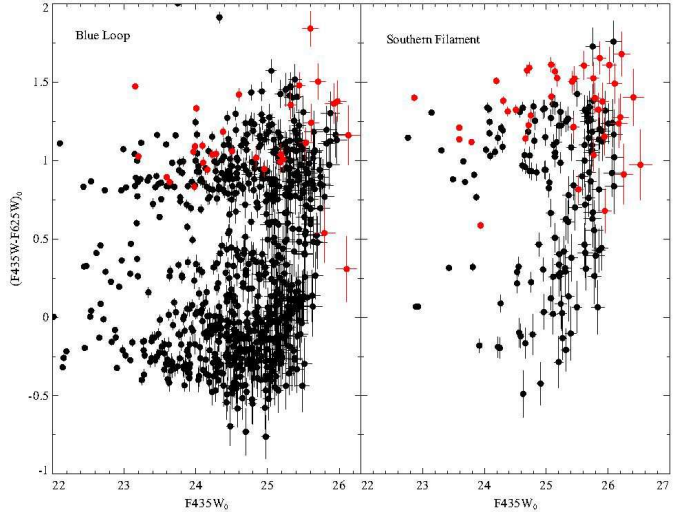}
\caption{Colour-magnitude diagram showing a comparison of the results for the Blue Loop and Southern Filament regions (in black) and for two control regions (in red) of size 15''$\times$15'' taken next to these regions.   }
\end{figure}

{\sc IRAF} {\sc PHOT} was employed to perform aperture photometry using an aperture with a two-pixel radius in all regions
and all filters. The background in the Blue Loop ans Southern filament regions was estimated using the modal value in an annulus 10 pixels 
wide with an 
inner radius of 15 pixels. In all regions growth curves of the brightest, isolated star clusters in the field were 
constructed to test the background estimation. The galaxy background near the outer regions of star
formation, far from the nucleus, does not vary much with position, however the field here is crowded. The large inner 
radius of the background annulus insures we are sampling much more background than surrounding stars. In the central 
region an inner radius of 12 pixels and annulus 2 pixels wide, as in \cite{carlson1998}, was found sufficient to estimate 
the background. The small sky radius here is used to minimise the variation of the background due to the galaxy. No 
subtraction of the smooth galaxy light in the inner region was done as the error introduced in the bright sources of 
interest is small. The worst case scenario of a non-linear background variation over the 28 pixels, for the innermost 
and therefore most steeply varying background introduces an error of less than 0.3 counts~s$^{-1}$. The variation of 
background to peak for this innermost cluster is less than 2 per cent. In the inner region the bright galaxy nucleus 
was masked out, as was a bright saturated star near the top of the eastern arm of the Blue Loop.

Average aperture corrections are determined using the 10 brightest sources in each of the three regions. 
These are corrections from a two pixel radius aperture corresponding to 
a 0.1 arc-second radius aperture flux to a flux determined with a 0.6 arc-second radius.  
These corrections are listed in Table 1.  The corrections are similar but slightly larger than
those found by \cite{carlson1998} in the core of the galaxy. The aperture correction for the three stellar regions was 
determined separately and is found to be marginally larger in the outer regions. The point spread function changes with 
position on the chip (errors of a few per cent), due to both optical aberrations and geometric distortions which may be 
responsible for the small differences detected in aperture correction across the field.

Fig. 5 shows a colour-magnitude plot of the outer star formation regions accompanied by photometry of neighbouring control regions. We 
do not find any objects as blue as our blue population of clusters in any of these fields, however the red population 
appears ubiquitous throughout the field. Contamination of the young star cluster population from foreground and background 
sources is therefore highly improbable due to both the lack of young blue sources in other regions and the spatial 
distribution of the blue population in the Blue Loop and Southern filament regions. This is discussed further in section 3.1 and Fig. 7.

The ACS filters differ significantly from other filter systems and this needs
to be taken into account when determining reddening corrections from the galactic foreground.
Due to the differences in filter transmission curves between ground-based and the ACS filter systems
the extinction coefficients are calculated in the native photometric system and
all corrections applied before converting to another system.

For comparison with the stellar evolutionary synthesis models, we transform the ACS filter system to the Johnson-Cousins UBVRI system using the method below,
described in \cite{holtzman1995} and \cite{sirianni2005},
\begin{equation}
TMAG=SMAG+c0+c1 \times TCOL+c2 \times TCOL^{2}.
\end{equation}
The F435W filter can be transformed to a B band filter and F625W to a R band filter,
however uncertainties of a few per cent are introduced during this transformation. An additional systematic error is introduced when using the synthetic 
transformation coefficients for stars with B$-$V$<0.5$ due to uncertainties in the shape of the total response curve in the F435W filter 
(see \citealt{sirianni2005}, their Fig. 21). The overall uncertainty in these transformations is of the level 5-10 per cent.
For comparison Fig. 6
shows the B-band magnitude verses colour diagram for the 
clusters after conversion to the Johnson-Cousins system. We used the synthetic transformation
coefficients given in Appendix D Table 22 in \cite{sirianni2005}
as these cover the larger colour range necessary for these results. We required the convergence
criterion to be $1\times10^{-5}$.

\begin{figure}
\centering
\includegraphics[width=0.45\textwidth]{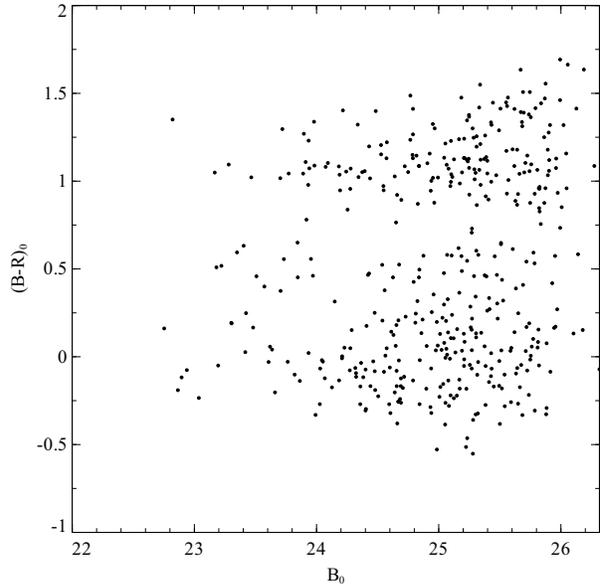}
\caption{Johnson-Cousins $(B-R)_{0}$ colour against B-band Vega magnitude diagram for the Blue Loop region.} 
\end{figure}

The extinction coefficients are determined by a constant times the reddening,
\begin{equation}
\frac{A_{\lambda}}{E(\lambda-V)}=R_{\lambda}.
\end{equation}
Here we have assumed
the constant, the ratio of total to selective extinction, is $R_{B}=3.1$ \citep{pei1992}.
This is the standard value derived from the $(B-V)_{0}$ colours in the Johnson-Cousins UBVRI band-passes,
using a ground based system. 
The extinction ratios in the ACS filters were derived by \cite{sirianni2005} using the 
\cite{cardelli1989} Galactic extinction law. The throughput in the medium and broadband
filters depends on the colour of the object. \cite{sirianni2005} selected three template
stars (05 V, G2 V and M0 V) from the Bruzual-Persson-Gunn-Stryker (BPGS) atlas of which we have adopted 
the 05 V results to compute the extinction coefficients in the three filters (Table 2). The results are
normalised to the value $R_{B}=3.1$, however, there are systematic differences in extinction
between the ACS system and a ground based system as a function of $E(B-V)$ due to differences
in effective wavelength. The extinction in our ACS wide-band filters is systematically higher than 
in the U B and V bands
as the effective wavelength is shorter. In the F814W 
band the effective wavelength is longer and so the extinction here is systematically lower than in the I band \citep{sirianni2005}. 
The value for reddening in the
direction of \n\ we use is $E(B-V)=0.1627{\pm}0.0014$ determined using the Galactic reddening maps of \cite{schlegel1998}.

There will also be reddening from the dust in NGC 1275 itself. This will cause a greater
effect towards the centre of the galaxy, especially near the HVS, resulting 
in the measurement of redder colours. The internal extinction is difficult to constrain, however
the two main areas of interest in this work, the Blue Loop and the Southern filament, are far from the dust lanes \citep{keel2001},
so we have not attempted to add a correction for this effect.

\section{Analysis and Results}

The H$\alpha$ filaments belonging to both the LVS and HVS have been investigated by \cite{conselice2001} (their Fig. 2.), and  \cite{caulet1992} (their Fig. 2.) respectively. Their images show the extended filamentary system is predominantly associated with the large central galaxy not the smaller high velocity galaxy in the line of sight. \cite{hatch2006} identifies a 1-arcmin-long chain of blue star clusters in the north-west of NGC 1275 (RA $03^{h}19^{m}46^{s}.7$, Dec. $41^{\circ}$ 31'45.6'' J2000), an extension of those mentioned by \cite{conselice2001}. Only one blue knot within this chain appears to be spatially connected with the H$\alpha$ filaments. \cite{hatch2006} find that this blue knot has a line-of-sight velocity of 5538$\,$kms$^{-1}$ placing it within the LVS. 

The following analysis assumes the star clusters along the two extended regions of star formation studied here also belong to the LVS (see also later discussion at end of section 4 and Fig. 14). Future spectroscopic data of these star clusters would allow us to place the clusters firmly within the HVS or LVS.
\begin{figure}
\centering
\includegraphics[width=0.5\textwidth]{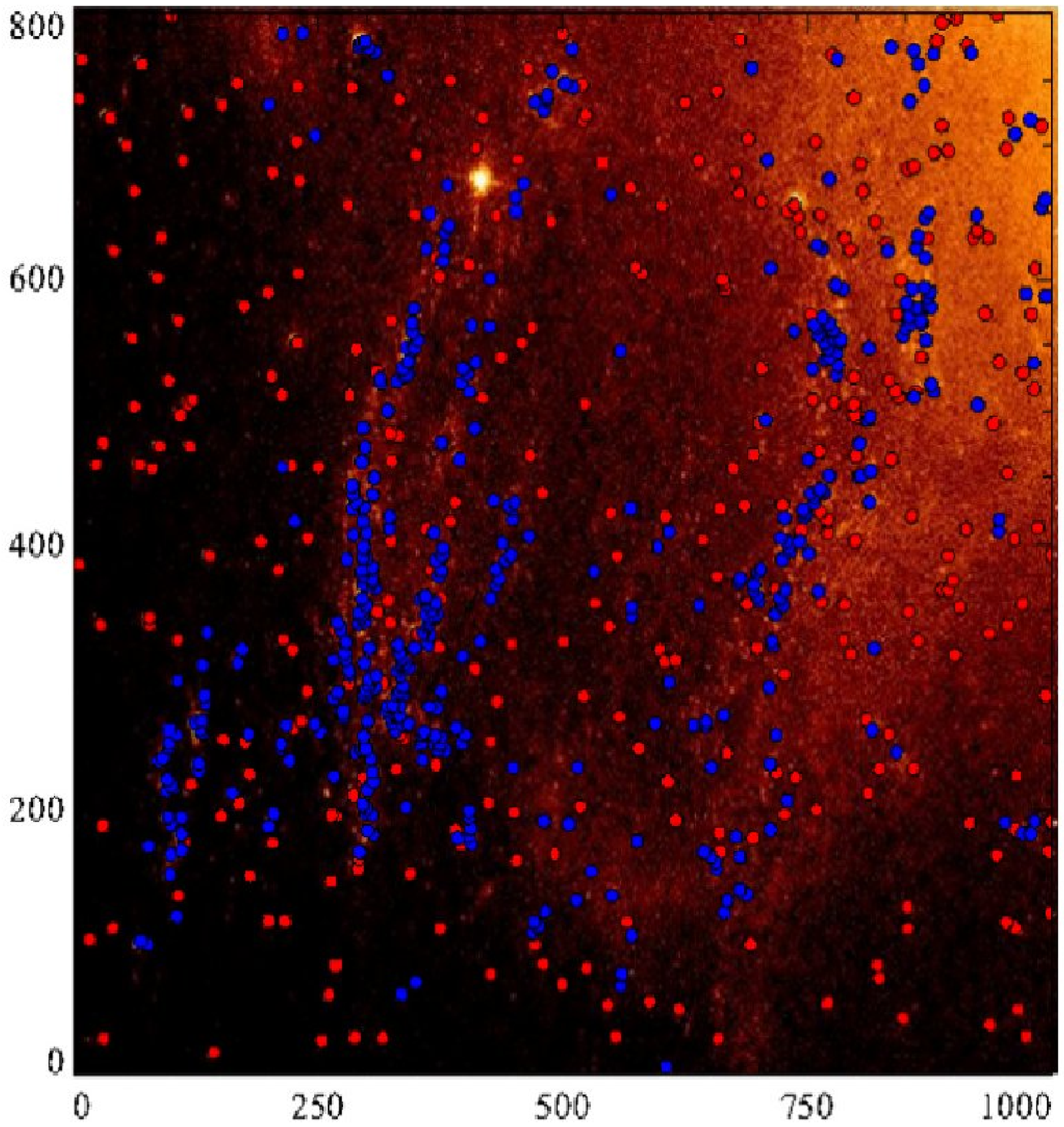}
\includegraphics[width=0.5\textwidth]{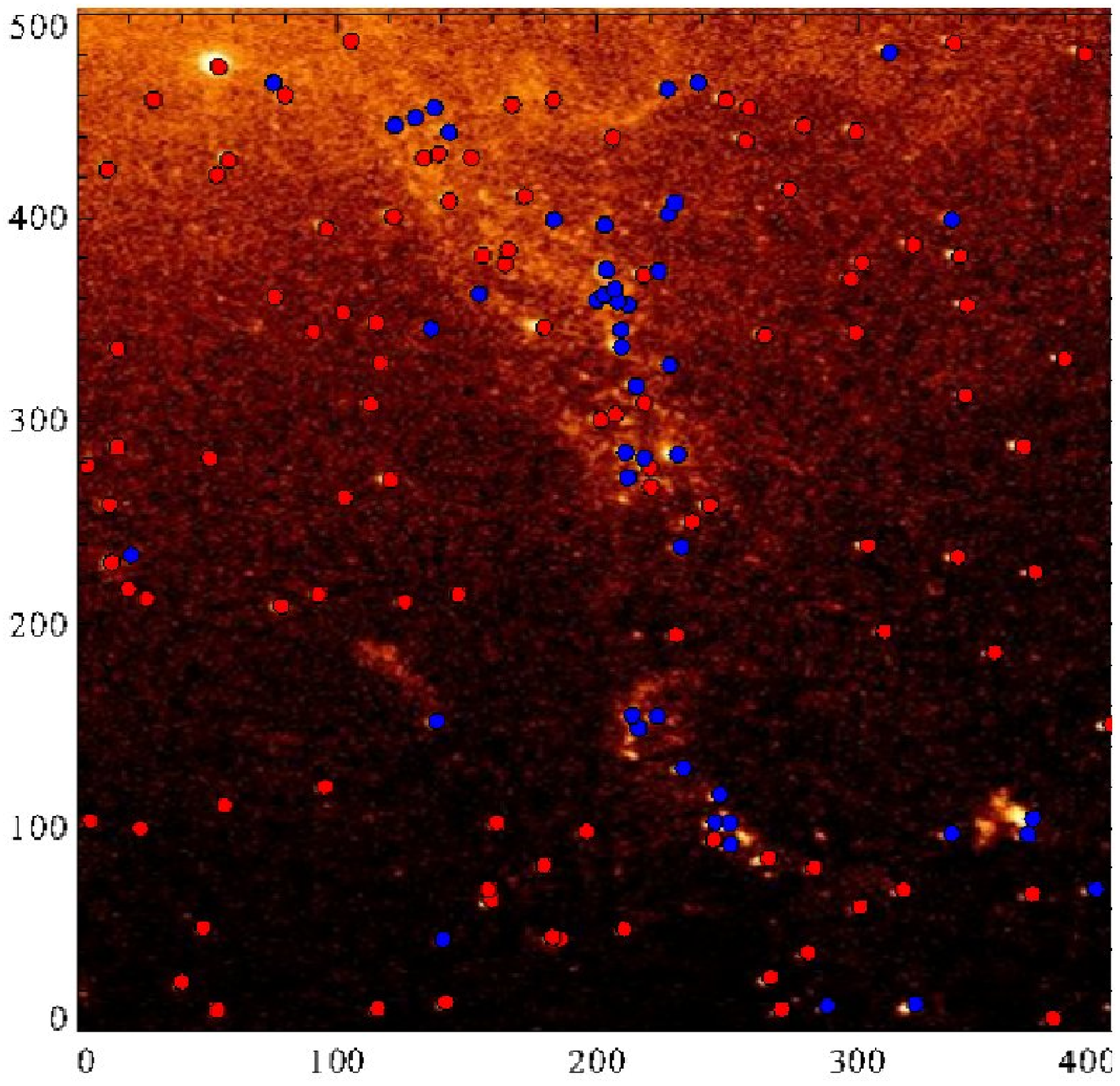}
\caption{Images showing spatial arrangement of blue and red clusters in both the Blue Loop (upper) and the Southern filament (lower). $x$ and $y$ axis here are both in pixels. In the Blue Loop region blue clusters are defined as those
with de-reddened $(\mathrm{F435W}-\mathrm{F625W})_{0} < 0.4$ and red clusters with de-reddened $(\mathrm{F435W}-\mathrm{F625W})_{0} > 0.4$. In the Southern filament blue clusters have $(\mathrm{F435W}-\mathrm{F625W})_{0} < 0.5$ and red clusters with $(\mathrm{F435W}-\mathrm{F625W})_{0} > 0.5$. These populations are determined by inspection of the colour histogram (see Fig. 4) in each region.}
\end{figure}
\subsection{Spatial Distribution of Clusters}

The clusters in the south-west region contain a range of colours evenly distributed about the core of the galaxy, implying the old and young population are inter-dispersed throughout the core of the LVS. The south-west region is spatially the most distinct region from the HVS in the core of NGC 1275, however confusion with this system can not be ruled out.

In the Blue Loop region all `bluer' clusters (defined as those with de-reddened $(\mathrm{F435W}-\mathrm{F625W})_{0} < 0.4$) are found to lie along the loop, which coincides with the H$\alpha$ emission (see Fig. 7). The `redder' clusters (de-reddened $(\mathrm{F435W}-\mathrm{F625W})_{0} > 0.4$) are evenly distributed and have the same colours as clusters found in a control region of the galaxy at a similar radius. The control region used has a central RA $03^{h}19^{m}54^{s}.1$ and Dec. $41^{\circ}$ 30'37.7'' (J2000) and was 10 by 10 arc-seconds across. This is also found to be the case in the Southern filament region, here blue cluster have $(\mathrm{F435W}-\mathrm{F625W})_{0} < 0.5$ and red clusters have $(\mathrm{F435W}-\mathrm{F625W})_{0} > 0.5$. Projection effects make it difficult to be precise about the arrangement of the extended filamentary systems in NGC 1275, however the close nature of the filaments and the stellar regions, and the fact that we see this apparent association in more than one area, yet the filaments themselves are very sparse in the extended regions, suggests that the star formation is likely to be intimately linked to the optically-emitting filaments.

\subsection{Stellar Populations Models}
We use Simple Stellar Populations (SSP) models to estimate the ages of the young populations of clusters. It is likely that these clusters formed almost simultaneously and with similar metallicities allowing them to be described by these models.

The commonly used models of \cite{bc03} (hereafter BC03) and the GALEV evolutionary synthesis models \citep{Kotulla2009} have been used to determine an upper bound on the ages of the star clusters
in NGC 1275. The GALEV models have the advantage of including the ACS filter system while the BC03 models require a photometric transformation to the Johnson-Cousins UBVRI system. This transformation introduces errors of a few per cent \citep{sirianni2005}.
\begin{figure}
\centering
\includegraphics[width=0.45\textwidth]{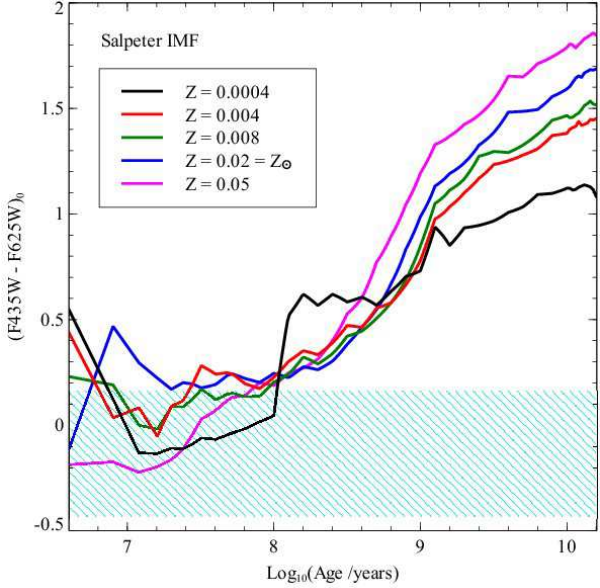}
\caption{($\mathrm{F435W}-\mathrm{F625W})_{0}$ colour verses age relationship for several metallicities, measured relative to solar, using the GALEV SSP models, assuming a Salpeter IMF with lower mass cut-off $m_{L}=0.1\,\mathrm{M}_{\odot}$. The upper mass cut-off depends on metallicity with $m_{U}=50\,\mathrm{M}_{\odot}$ for super-solar and $m_{U}=70\,\mathrm{M}_{\odot}$ otherwise. The hashed lines show the spread in colour of the young, blue population in the Blue Loop region.} 
  \end{figure}
  \begin{figure}
\centering
\includegraphics[width=0.45\textwidth]{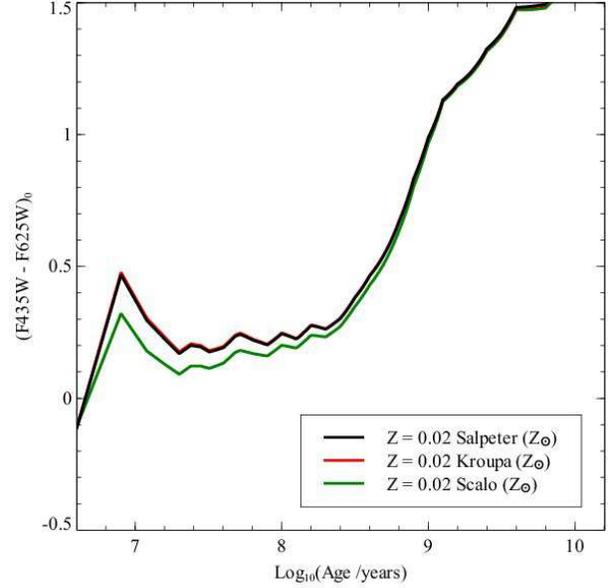}
\caption{The $(\mathrm{F435W}-\mathrm{F625W})_{0}$ colour verses age relationship using Galev SSP models assuming solar metallicity with a Kroupa, Salpeter and Scalo IMF respectively.}
  \end{figure}
  \begin{figure}
\centering
\includegraphics[width=0.45\textwidth]{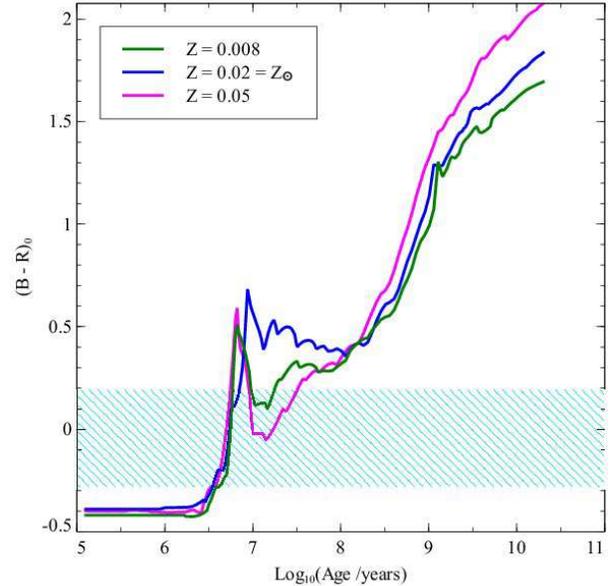}
\caption{Johnson-Cousins $(B-R)_{0}$ colour versus age relationships for three different metallicities using BC03 models for a single age, single burst population and assuming a Salpeter IMF. Lower and upper mass cut-offs are assumed to be $m_{L}=0.1\,\mathrm{M}_{\odot}$ and $m_{U}=100\,\mathrm{M}_{\odot}$ respectively. The hashed lines show the spread in colour of the young, blue population in the Blue Loop region.}
  \end{figure}
The GALEV SSP models are shown for a range of metallicities and Initial Mass Functions (IMF's) ($\Phi(m)^{\alpha}$) in Fig. 8 and 9. These are all computed using the Padova 1994 isochrone models. \cite{Kotulla2009} assume a lower mass cut-off of $m_{L}=0.1\,\mathrm{M}_{\odot}$ while they allow the upper mass cut-off to depend on metallicity. A cut-off of $m_{U}=50\,\mathrm{M}_{\odot}$ is assumed for super-solar metallicity and $m_{U}=70\,\mathrm{M}_{\odot}$ otherwise. The Salpeter, Scalo and Kroupa IMF's assumed are defined with $\alpha$ given in equations 3, 4 and 5 respectively.

\begin{equation}
\alpha=-2.35
\end{equation}
\begin{equation}
\alpha \left\{
 \begin{array}{rl}
-1.25 & \text{if } m \leq 1\,\mathrm{M}_{\odot}\\
-2.35 & \text{if } 1\,\mathrm{M}_{\odot} < m \leq 2\,\mathrm{M}_{\odot}\\ 
-1.25 & \text{if } 0.5\,\mathrm{M}_{\odot} < m
 \end{array} \right.
\end{equation}
\begin{equation}
\alpha \left\{
 \begin{array}{rl}
-1.30 & \text{if } m \leq 0.5\,\mathrm{M}_{\odot}\\
-2.30 & \text{if } 0.5\,\mathrm{M}_{\odot} < m 
 \end{array} \right.
\end{equation}
\\

The BC03 SSP models (Fig. 10) are computed assuming a Salpeter IMF
and the Padova 1994 evolutionary tracks. The models assume an IMF with a low mass cut-off 
of $m_{L}=0.1\,\mathrm{M}_{\odot}$ and a high mass cut-off of $m_{U}=100\,\mathrm{M}_{\odot}$. The models are normalised
to a total mass of $1\,\mathrm{M}_{\odot}$ in stars at age $t=0$.

X-ray observations have shown the metallicity of the ICM to be approximately 0.6 of solar abundances at the Blue Loop region \citep{sanders2004, sanders2007}. Given this, metallicity abundances as high as twice solar or higher (Z$=0.05$), in these stars, are very unlikely as are abundances as low as 2 per cent (Z=0.0004) solar (Fig. 8). We might expect differences in the metallicities of the old and young cluster populations, the old clusters are potentially metal-poor globulars while the young clusters are likely to have metallicities $\sim$ solar. The SSP models used agree well for ages larger than 10$^{7}$ years, however modelling very young stars is complex and the models remain uncertain. This, coupled with errors on the data make it impossible to constrain the metallicities of the regions.

The observed colour of the clusters are highly sensitive to age, metallicity, dust extinction and the coincidence of the clusters with H$\alpha$ filaments. Older, metal-rich, dusty populations will appear redder. The star formation in regions such as the Blue Loop is far from the nucleus ($\sim22\,\mathrm{kpc}$), and exhibits no obvious signs of dust, so internal reddening is unlikely to have a large effect. As discussed above the metallicities are likely to be $\sim$ solar, however, as seen by Figs. 8 and 10, changes in metallicity can have a significant effect on the colour of young stellar populations. The clusters in our sample are coincident with H$\alpha$ emission. This will have the effect of reddening the colours of these clusters and may make them appear older than they are if the emission is within the photometry aperture or make them appear bluer if the emission is within the background annulus. There are ~twice as many counts in the brightest filamentary regions on the Blue Loop than counts in the background and for fairly bright sources (25 mag) the background counts are only ~2 per cent of the counts from the cluster. To make a 10 per cent difference to the $(\mathrm{F345W}-\mathrm{F625W})_{0}$ colour the H$\alpha$ (+[NII]) emission would require an equivalent width of $>100\,\AA$. This is very large and unlikely to be an issue.

\begin{figure}
\centering
\includegraphics[width=0.5\textwidth]{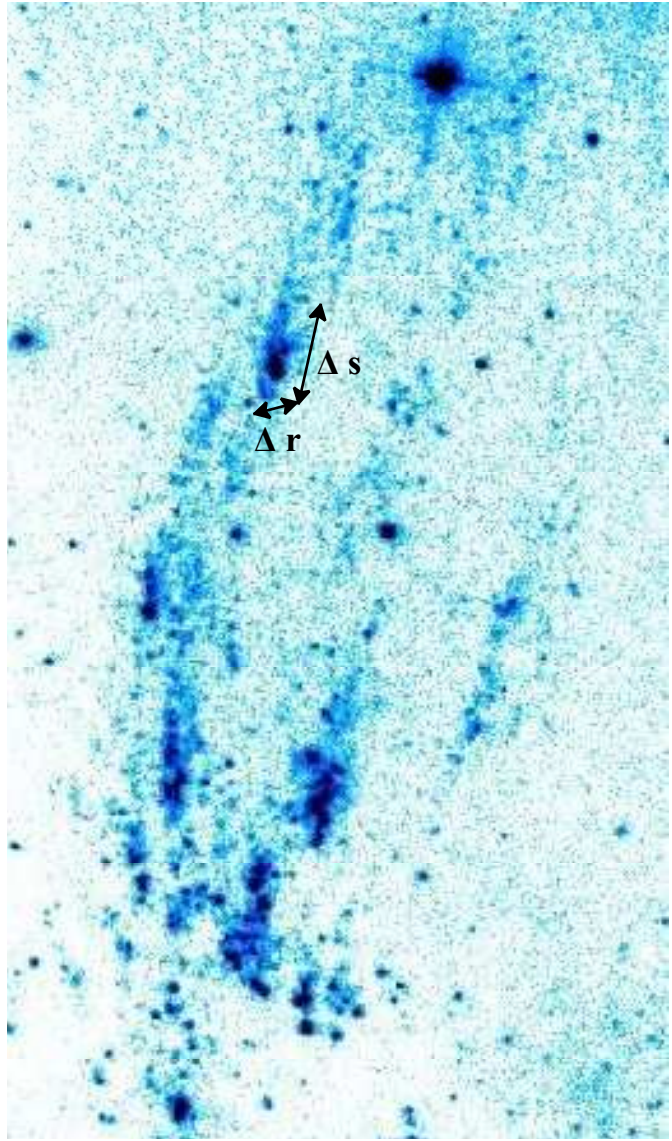}
\caption{The eastern arm of the Blue Loop showing that the clusters appear elongated possibly indicative of tidal stripping by the galaxy core.}
\end{figure}
\subsubsection{Blue Loop}
The bimodal distribution in colour shown in Fig. 4 strongly argues for a two-age population of clusters in this region. The colours of the `bluer' region are very blue with an average, extinction corrected colour of $(\mathrm{F435W}-\mathrm{F625W})_{0} = -0.2$, much bluer than the proto-globular clusters found in the core of the system ($\sim 0.3$). This also argues for a low internal extinction in this region. 

Assuming a metallicity of $Z=0.02$ ($=Z_{\odot}$) for the star clusters and a colour range between $-0.4$ and $0.2$, 
lower and upper age estimates from the BC03 SSP models are $5\times10^{6}$ and $5\times10^{7}$ years. The GALEV models yield an upper limit of $0.75\times10^{8}$ years.
The ages and luminosities can also give us estimates of the total mass in the clusters and hence
put limits on the star formation rates. We get a lower limit of $9\times10^{8}\,\mathrm{M}_{\odot}$
for all the blue clusters summed together. This provides a limit on the star formation of about
20 $\mathrm{M}_{\odot}\,\mathrm{yr}^{-1}$ over the whole Blue Loop region. The average mass in the clusters is found to be $4\times10^{6}\,\mathrm{M}_{\odot}$ with the brightest clusters having masses $\sim10^{7}\,\mathrm{M}_{\odot}$.

In the central region, \cite{carlson1998} found mass estimates for their brightest clusters of about $10^{7}-10^{8}\,\mathrm{M}_{\odot}$.
These results are highly dependent on the IMF and the models assume
a single burst, single age population. In the Blue Loop, the relatively large amount of scatter in panel (a)
in Fig. 3 suggests, assuming these clusters all share the same metallicity, that it does not have a true single age population. This could be evidence of continuous star 
formation which `turned on' approximately $10^{8}$ years ago, the likely dynamical lifetime of the H$\alpha$ filaments \citep{hatch2005}. However we note, for very young stellar populations ($<10^{7}$~years) the SSP models are uncertain and the breadth of the colour population may not necessarily translate into a large age difference. A stellar population with an age 10$^{6}-10^{7}$~years can span the whole range of observed colours using BC03 models while the GALEV SSP models cannot account for any of the bluest clusters observed.

There is also a large amount of diffuse blue light around the Blue Loop region and particularly 
on the north-west side towards the galaxy. This is most likely emission from smaller clusters and
single stars, possibly tidally stripped, or dissolved from, surrounding larger clusters. If these stars are similar in age to the clusters in the loop we would expect the total mass in stars, and therefore also the star formation rate calculated, to be a lower limit.

\subsubsection{Southern Filament}
The Southern filament region shows slightly redder colours, $(\mathrm{F435W}-\mathrm{F625W})_{0}$ $\sim$ $-0.1$ to 0.2, than the Blue Loop region, $(\mathrm{F435W}-\mathrm{F625W})_{0}$ $\sim$ $-0.4$ to 0.1 with a lower and upper 
limit on the ages of the star clusters of about $8\times10^{6}$ and $5\times10^{8}$ years from the BC03 models and an upper limit of $5\times10^{8}$ years from the GALEV models. Fig. 3 panel (b) shows there is a clear bimodal distribution in this region however the distribution of colours in the young blue population is again fairly broad.

\subsubsection{South-west Nuclear Region}
The age estimates from the models applied to the south-west portion of the nucleus give limits
of $10^{7}$ and $10^{9}$ years, in agreement with the values estimated by \cite{carlson1998} and with the spectroscopically determined ages of five of the brightest inner clusters by \cite{brodie1998}.
\begin{table}
\begin{center}
\begin{tabular}{|c|c|c|c|}
\hline
ID & F435W$_{0}$ & F625W$_{0}$ & $(\mathrm{F435W}-\mathrm{F625W})_{0}$ \\
\hline \hline
1  & 21.559$\pm$0.013 & 21.006$\pm$0.011 & $\,0.55\pm0.017$ \\
2  & 21.324$\pm$0.011 & 21.349$\pm$0.014 & $-0.03\pm0.018$ \\
3  & 21.801$\pm$0.016 & 21.697$\pm$0.019 & $\,0.10\pm0.025$ \\
4  & 21.083$\pm$0.008 & 21.088$\pm$0.011 & $-0.00\pm0.014$ \\
5  & 21.183$\pm$0.010 & 21.255$\pm$0.014 & $-0.07\pm0.017$ \\
6  & 21.220$\pm$0.010 & 21.218$\pm$0.013 & $\,0.00\pm0.016$ \\
7  & 21.992$\pm$0.025 & 21.958$\pm$0.032 & $\,0.03\pm0.041$ \\
8  & 21.393$\pm$0.014 & 21.439$\pm$0.018 & $-0.05\pm0.023$ \\
9  & 21.707$\pm$0.013 & 21.700$\pm$0.016 & $\,0.01\pm0.021$ \\
10 & 20.898$\pm$0.006 & 21.006$\pm$0.009 & $-0.11\pm0.011$ \\
\hline
\end{tabular}
\end{center}
\caption{Magnitudes and colours for apertures 1 to 10 shown in Fig. 12.}
\end{table}

\subsection{Tidal Forces}

Fig. 11 shows that there are `groups' of star clusters in the Blue Loop which have a structure 
reminiscent of a tail. One explanation for this could be that when the stars have formed they are unable to escape the gravitational potential of the galaxy and fall
back in towards the the nucleus.

If the star clusters have been tidally stripped, we would expect a gradient in the ages of the blue population. The star clusters forming first would have been subject to tidal stripping for longer and so should have fallen further towards the galaxy.

\begin{figure}
\centering
\includegraphics[width=0.5\textwidth]{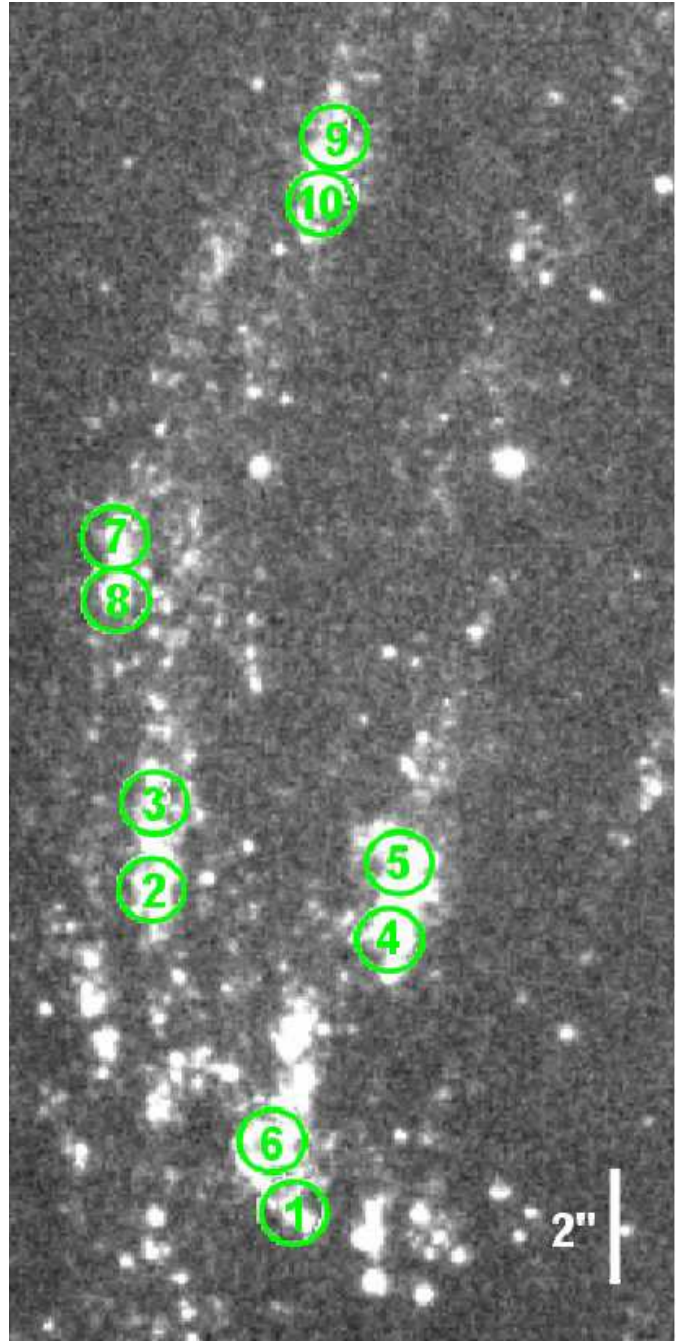}
\caption{Apertures with a 9 pixel radius to test whether there is any discernible age difference along the `groups' of star clusters. The results of the photometry are given in Table 3.}
\end{figure}
To detect whether there is a discernible age difference across the star cluster `groups' in the bright eastern arm of the Blue Loop we choose 5 prominent groups of clusters. These are shown in Fig. 12 and the results of the photometry are given in Table 3. Three of the regions, those with apertures 2/3, 8/7 and 10/9 are found to have the younger, bluer clusters towards the south and two, 1/6 and 4/5, the younger clusters to the north. The difference detected in both 8/7 and 4/5 is small compared with the error in colour. The detected difference in colour in region 1/6 is very large but on closer inspection of the apertures it is not clear that they cover the same group of clusters so this region should be discounted.

Region 2/3 and 10/9 show a difference in colour larger than the error (change of 0.4 and 0.12 between apertures respectively). In both regions the younger clusters appear to lie in the southern most aperture. This is expected if the star clusters have formed at different times and not in some rapid event.

The age of the clusters can be estimated by assuming each elongated group to be experiencing
tidal stripping as it falls into the nucleus. The groups have an initial angular momentum from
the group formation process. The ratio of the length to width ($\Delta s / \Delta r \approx 5$, see Fig. 11) of each of these groups 
of clusters can then be measured and the time over which they have been
stripped equated with the age by making the assumption that as soon as the gas was able to condense
out of the filament and form stars they started to fall into the nucleus. The typical ratio of $\Delta s$ to $\Delta r$ is measured from the ACS images note though, that projection effects could significantly alter this ratio.

Equation 6 and 7 assume that the clusters within each of these groups started to form in one place so the spread of the group ($\Delta s$ and $\Delta r$) seen now is the result of the initial velocity of the gas and tidal forces from the attraction of the LVS.
The gravitational acceleration felt by the element is $g\approx v^{2}/r$ where $v=700\,\mathrm{km}\,\mathrm{s}^{-1}$ is the velocity
dispersion. The velocity dispersion is inferred from the temperature of the ICM \citep{fabian2006} measured in X-rays. The tidal acceleration is given by,
\begin{equation}
 \Delta g \approx \frac{GM \Delta r}{r^{3}} \sim \frac{g\Delta r}{r}
\end{equation}
so after time $t$,
\begin{equation}
 \frac{\Delta s}{\Delta r} = \frac{1}{2}\frac{v^{2}t^{2}}{r^{2}}
\end{equation}
The distance from the centre of the potential to the group is $r = 63''$ which is about $22\,\mathrm{kpc}$ at the 
distance of NGC 1275. With the ratio $\Delta s/\Delta r \approx 5$ and a velocity dispersion of $v=700\,\mathrm{km}\,\mathrm{s}^{-1}$ we calculate an age of $9\times10^{7}$ years, approximately the same as the 
upper bound on the estimate from the SSP models. The precise nature of the tidal distortion for either radial or tangential motion depends on $v(r)$ which in turn depends on the mass distribution.

\section{Discussion}

In this paper we present photometry results for star clusters close to the nucleus and at a radius of $\sim$ 1 arcmin in the dominant galaxy in the Perseus cluster, NGC 1275. We give limits on the ages and masses of the star clusters and find a marked difference in the ages of the stars in the outer regions compared to those closer to the nucleus of the galaxy. In this section we discuss the implications of these results for the formations scenarios of the inner and outer cluster regions in turn.

\subsection{Inner Stellar Regions}

Our results, corrected for galactic extinction but uncorrected for internal extinction, yield colours of $(\mathrm{F435W}-\mathrm{F625W})_{0}$ $\sim$ 0.2 to 0.6 for the bluer star cluster population in the centre of NGC 1275. From SSP models we derive lower and upper limits on the ages of the clusters as $10^{7}-10^{9}$ years respectively, and an upper limit on the mass of the brightest clusters of $10^{8}\,\mathrm{M}_{\odot}$.

\cite{holtzman1992} were the first to discuss the formation scenarios of the bimodal population of star clusters in the LVS in NGC 1275. They detected a central population of blue clusters and explored three possible formation processes: (1) An interaction with the HVS, (2) star formation triggered by the cluster cooling flow and (3) star formation from a previous interaction with a third gas-rich galaxy. They rule out (1) primarily due to the high velocity ($3000\,\mathrm{km}\,\mathrm{s}^{-1}$) at which the two systems are falling toward each other and the clusters being roughly symmetrically distributed around the nucleus, rather than having a preferential direction towards the HVS. This conclusion was also reached by \cite{unger1990}. The second scenario was ruled out by the authors on the basis of the observed uniformity of the colour of the young, blue population of clusters. The same grounds favour scenario (3). This latter interpretation is supported by HST WFPC2 observations made by \cite{carlson1998}. However, \cite{richer1993} found a broader range in colours with no obvious bi-modality in their colour-magnitude diagram, which could indicate a large range in cluster ages and favours a scenario whereby the star formation is a formed in a more continuous manner. This is supported spectrally by the work of \cite{ferruit1994} who found the gas coincident with some of the inner clusters to have spectral properties similar to the gas in the filaments.

Two issues have changed since the earlier work involving scenario (3). The first is that the apparent destruction of the HVS at over 100$\,\mathrm{kpc}$ out from NGC 1275 emphasises that the ICM can prevent gas-rich galaxies penetrating right to the centre. The second is the enormous abundance of molecular gas already at the centre (more than $5\times10^{10}\,\mathrm{M}_{\odot}$, \citealt{salome2006}) which can easily supply fuel for star formation if dragged out by bubbles.

\begin{figure}
\centering
  \includegraphics[width=0.5\textwidth]{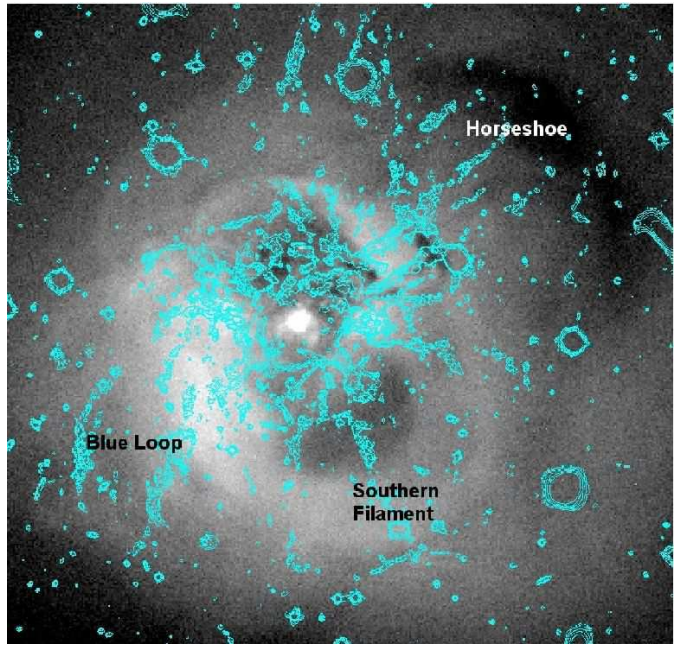}
\caption{X-ray image ($0.3 - 7.0\,\mathrm{keV}$) with overlaid H$\alpha$ contours.}
\end{figure}

In this work we find evidence for a bimodal distribution of colours in the star clusters located in the south-west portion of the nucleus. However we detect a larger scatter about the modal value than that found by \cite{carlson1998}. Difficulties in photometry caused by a highly varying background, uncertain internal extinction and confusion with the HVS, do not allow us to conclusively rule out a history of formation by an interaction/merger with another galaxy, or a history of extended star formation in the younger, bluer population of clusters in the centre of NGC 1275.

\subsection{Outer Stellar Regions}  
In both the Blue Loop and the Southern filament we find much `bluer' colours than in the core, with the former being the `bluest'. Colours in the Blue Loop of $(\mathrm{F435W}-\mathrm{F625W})_{0}$ $\sim$ $-0.4$ to 0.1 and the Southern filament of $(\mathrm{F435W}-\mathrm{F625W})_{0}$ $\sim$ $-0.1$ to 0.2, yield upper limits on the age of the extended star formation of $5\times10^{7}$ and $10^{8}$ years respectively, a factor of 10 younger than the blue clusters in the central region. 

\cite{conselice2001} found colours on the bright eastern wing of the Blue Loop to be in the range $(B-R)_{0}$ $\sim$ $-0.3$ to 0.0 consistent with our results. The very blue colours and the large distance from the nucleus ($\sim20\,\mathrm{kpc}$), found here suggests that the internal reddening in this region is small and argues strongly for a two-age population in these regions.

\subsection{Star Formation Scenarios}

We consider two formation scenarios for the outer stellar regions: (1) Formation related to a previous interaction assumed responsible for the inner stellar populations and (2) a formation scenario entirely independent of the young, inner stellar populations.

\subsubsection{(1) -- Star Formation Triggered by an Assumed Previous Interaction}

For star formation in the extended stellar regions to be triggered by a previous interaction, the young, blue populations should either be coeval or the interaction needs to have had a delayed effect causing the star formation in the inner regions to occur before the outer regions.

If the outer and inner regions of star formation are the same age there must be a large amount of internal extinction in the core of the galaxy. Assuming no internal extinction in the Blue Loop, the colour correction needed for this region to be of comparable age to the inner clusters is $E(B-V)$ $\sim$ 0.6. This means that the difference between the B and R band extinctions would need to be three times greater than the extinction calculated here to correct for Galactic reddening along the line of sight to NGC 1275. 
There is significant internal reddening from the LVS and also from the HVS. \cite{shields1990} use $A_{V}=1.2$ which gives an $E(B-V) = 0.39$ using the typical ratio of the extinction to the reddening of 3.1. We minimise the reddening due to the HVS by observing only the south-west, central region that is least obviously connected with the HVS.

The spatial distribution and distance from the core, of the extended star formation and the linear nature of the H$\alpha$ filaments argue against both of these scenarios. A merger or interaction of some kind is unlikely to disturb the Blue Loop and Southern filament systems enough to cause large amounts of localised star formation in these regions without disturbing other filaments nearby. The range of colours, especially in the Blue Loop, is also quite large, this could mean that the population is not all the same age and therefore unlikely to have been formed in some rapid event.

\subsubsection{(2) -- Formation Scenarios Independent of an Assumed Previous Interaction}

The extended blue populations of star clusters are particularly
intriguing as they appear spatially distributed along or just offset
from some radial H$\alpha$ filaments but only in certain regions of
the galaxy; {\it most H$\alpha$ filaments are not accompanied by star
  formation}. Another galaxy-galaxy
interaction is unlikely to be responsible for this young population for
both the reasons discussed above and due to the very young age of these
clusters, the oldest in the Blue Loop being at most $5\times$10$^{7}$~years.

With the HST images, \cite{fabian2008} have resolved
some filaments into a collection of threads, each of which has a
diameter of about 70~pc. They argue that magnetic fields close to the
pressure equipartition value of the hot gas, of about 100$\mu$G, are
required for the thin threads within the filaments to maintain their
integrity against tidal forces in the cluster core. Filaments with a
higher mean density would not be supportable. The general lack of
massive star formation within the filaments is then due to magnetic
fields which balance the self gravity of the cold gas (see
\citealt{mckee1993} for a review of the role of magnetic fields in
star formation and \citealt{ho2009} for a discussion on the inner
region of NGC\,1275).

\cite{fabian2008} find that a typical thread has a radius of 35~pc, a
length ($l$) of 6~kpc and a mass of about $10^6\mathrm{M}_{\odot}$ (obtained
by scaling the H$\alpha$ emission to a filament complex of total mass,
obtained from CO observations of $10^8\mathrm{M}_{\odot}$
\citealt{salome2008}). The mass of a whole 6~kpc long thread therefore
corresponds to the mass of one observed star cluster.  The
perpendicular column density $N\sim 4\times 10^{20}\pcmsq$ or
$\Sigma_{\rm \perp}\sim 7\times 10^{-4}\gpcmsq$. The lengthwise column
density, $\Sigma_{\parallel},$ is $\ell/2r$ times larger. The critical
surface density $\Sigma_{\rm c}$ for gravitational instability
corresponding to the magnetic field inferred in the filaments is given
by $\Sigma_{\rm c}=B/2\pi\sqrt{G}=0.062 (B/100\mu G)\gpcmsq$, which is
about two orders of magnitude larger than the inferred value for a
thread, which is therefore individually very stable against
gravitational collapse.

This leaves the question, why is there any star formation at all, and
why does it only occur in certain places?  We can only speculate here
but suspect that stars form when and where threads knot and clump
together to make much larger structures. There are many examples in
the images of NGC\,1275 where the filaments appear knotty. If a region
is disturbed or stretched in a radial direction, so that the cold gas can accumulate, then
clumps which are gravitational unstable can arise and form the star
clusters which have been the subject of this work. Examples where gas
may be accumulating as the field lines are stretched out are the ends
of the horseshoe (for this feature see Fig.~13 and Fig.~3 of
\citealt{fabian2008}). There appears to be no star formation there at
present but H$\alpha$-bright knots are seen. 

Clumps or parts of filaments which are gravitationally unstable have a
column density which is much too high to be ``pinned'' to the
surrounding hot gas by magnetic fields and presumably have their own
ballistic orbits. The process of star formation might then be
dynamical. Threads within the long-lived outer filaments are supported
by magnetic fields and are stable to gravitational collapse. Where the
gas aggregates then they separate from the hot gas and, if sufficiently
dense, collapse to form star clusters.

\begin{figure*}
\centering
\begin{center}  \subfigure[]{\label{fig:edge-a}}\includegraphics[width=0.32\textwidth]{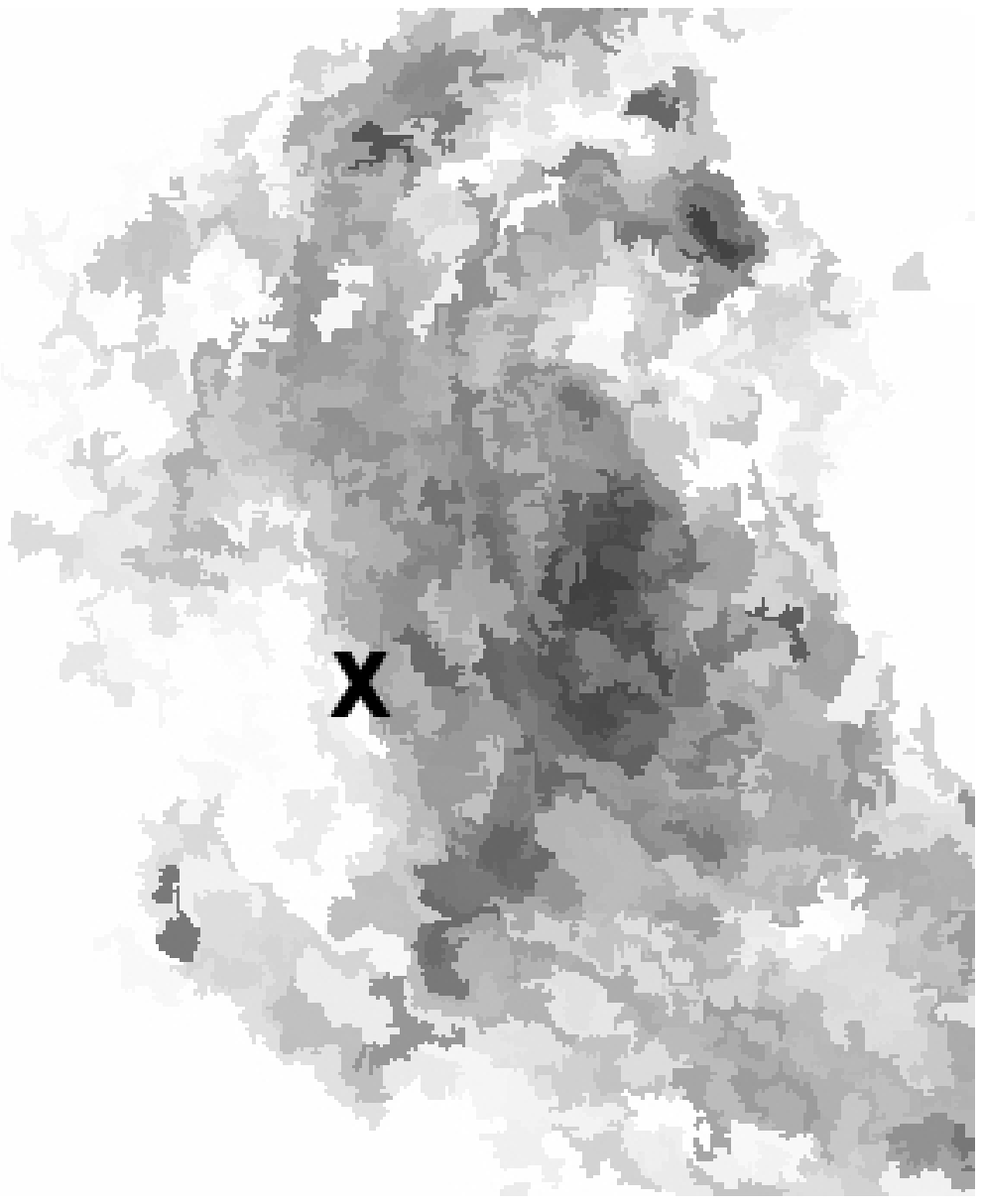}
  \subfigure[]{\label{fig:edge-b}}\includegraphics[width=0.32\textwidth]{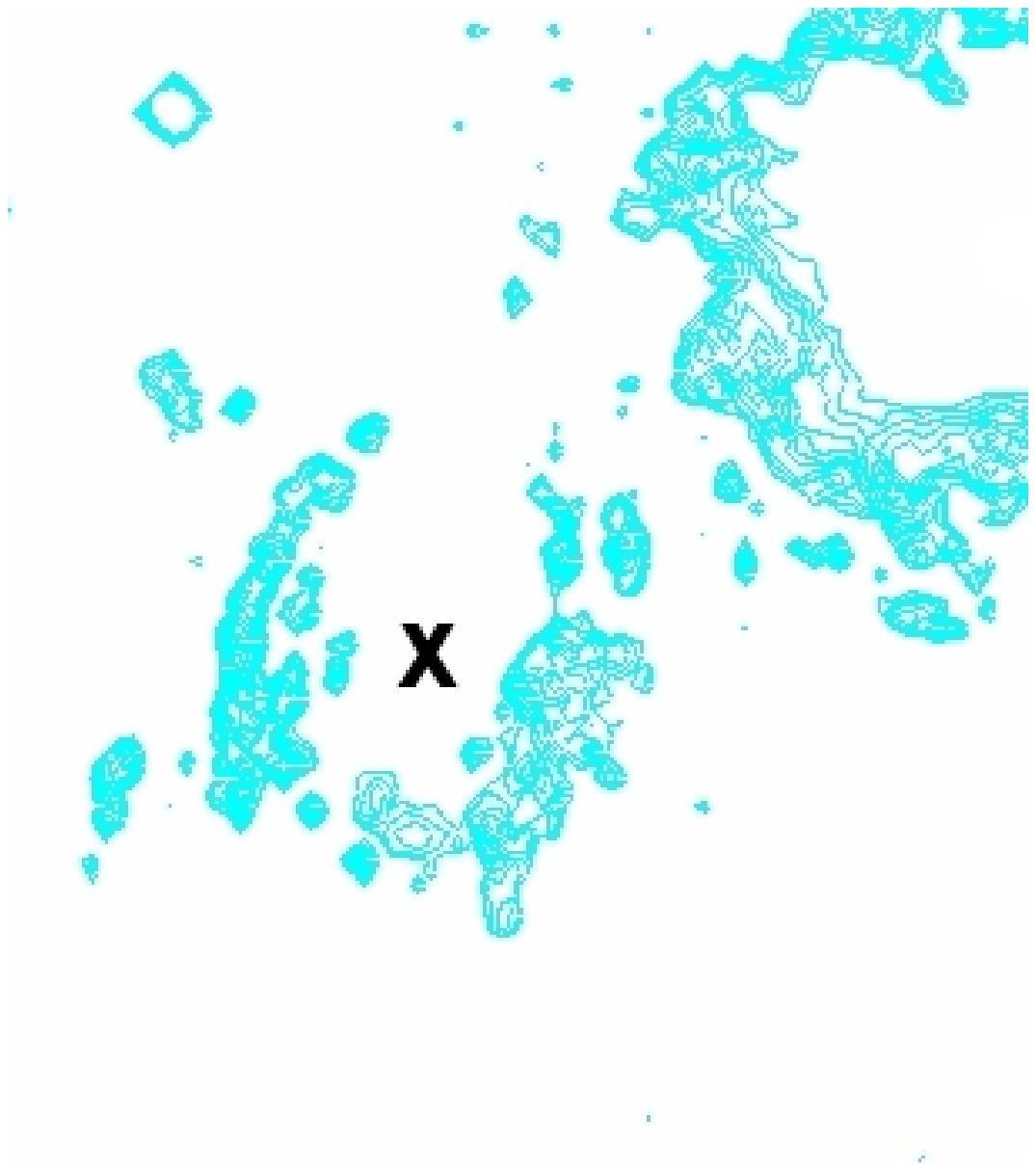}
  \subfigure[]{\label{fig:edge-c}}\includegraphics[width=0.32\textwidth]{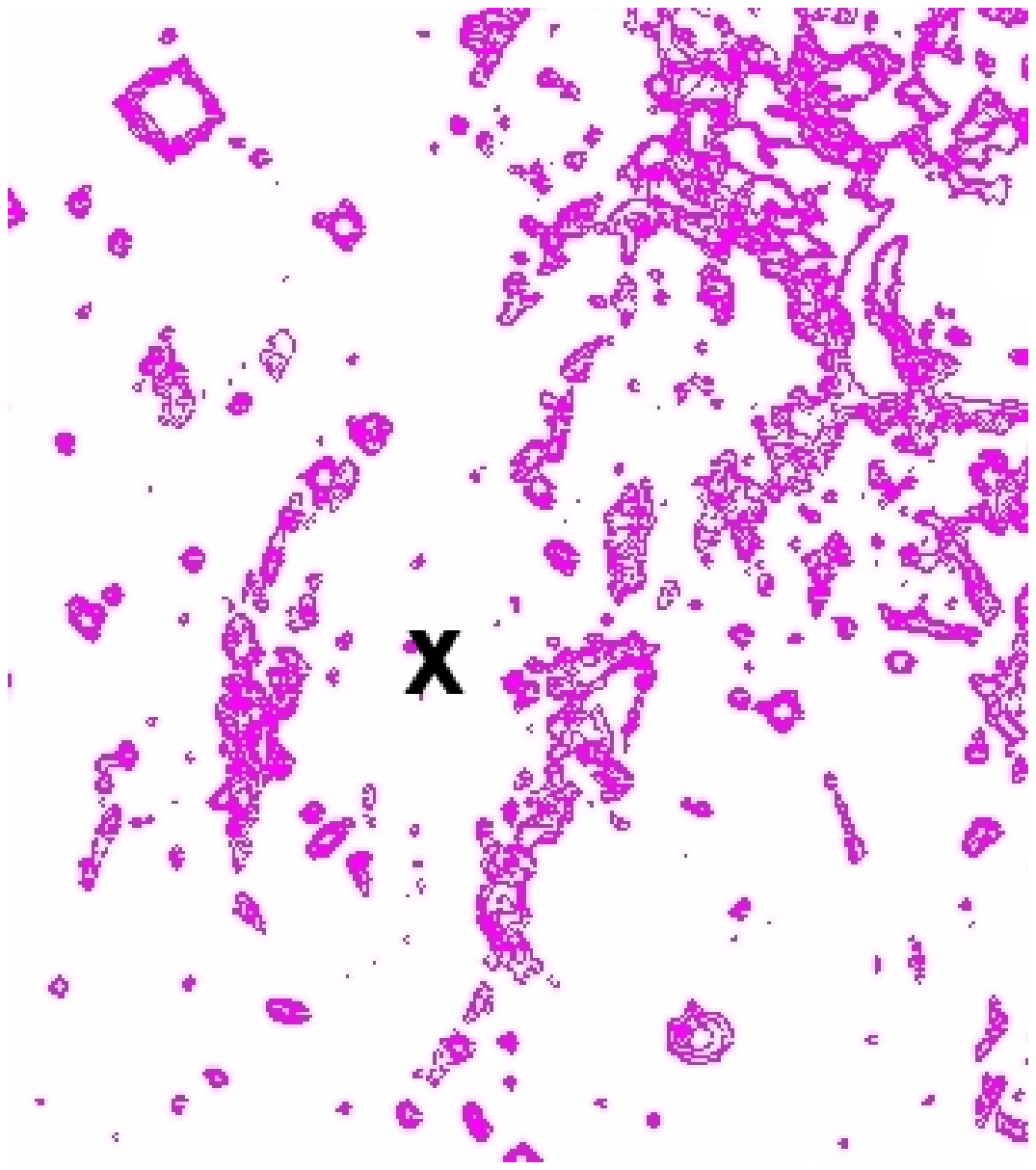}
\end{center}
\caption{The Blue Loop region: (a) X-ray temperature map, the dark regions are temperatures of $\sim2\,\mathrm{keV}$ and the light colours indicate temperatures of $\sim4\,\mathrm{keV}$. (b) B-band contours. (c) Contours of H$\alpha$ emission. The images are all the same scale, a black X marks the centre of the Blue Loop region on each image.}
\label{fig:edge}
\end{figure*}

We can put a limit on the maximum mass of a cluster using the
arguments of \cite{mckee1993}, assuming a steady, poloidal, time
independent magnetic field. Roughly, the magnetic energy scales as
$\mathcal{M}\sim B^{2}R^{3}$ and the gravitational energy scales as
$W\sim M^{2}/R$ so for a given field there is a critical mass where
the magnetic field can no longer support the filament against
gravitational collapse. This mass, $M_{\rm B}$, for a magnetised cloud
of radius $R$ and magnetic field $B$, with a constant of 0.12
determined from numerical calculations allowing for deviations from
uniform spherical clouds \citep{tomisaka1988}, is
\begin{equation}
 M_{\rm B}=0.12\times \frac{B R^2}{G^{1/2}}.
\end{equation}
Taking typical values for a thread, then $B R^2$ is
$\pi\times(35\,\mathrm{pc})^{2}\times100\,\mathrm{\mu G}$ resulting in
$M_{\rm B}$ at just under $10^{6}\,\mathrm{\mathrm{M}_{\odot}}$. This is just
less than the average mass of a young cluster in the Blue Loop region
of $4\times10^{6}\,\mathrm{M}_{\odot}$. In order for a thread-sized region to
achieve gravitational collapse its mean density needs to rise by about
x100 above what is inferred for a typical thread
(\citealt{fabian2008}). The cold material therefore needs to aggregate
considerably, as discussed above. Note that the mean density of a
resolved thread of $\sim 2\pcmcu$ is much lower than the actual
density of the cold gas ($\sim 100-1000\pcmcu$) if its thermal
pressure is comparable to the thermal pressure of the surrounding
X-ray emitting gas where $nT\sim 10^6\pcmcuK$. (in other words the
volume filling factor of cold gas in a thread is low.) 
    
The brightest star clusters have masses on the order
$10^{7}\,\mathrm{M}_{\odot}$. This requires them to originate from regions
several times larger than a thread. Since such regions cannot be
suspended by the hot gas, then the observed values of the thickness of
long-lived threads no longer apply. As
mentioned above one explanation is that the threads knot together and
collapse along their lengths, allowing them to evolve from stable to
unstable structures over time.

We can only speculate on what has triggered to star formation of the
Blue Loop. It coincides with a hotter region in the X-ray temperature
map (Fig.~14), which is the best evidence that the stars are
associated with the LVS and not the HVS. Clear confirmation must await
spectroscopic measurements of the star clusters. The blue loop roughly
lies diametrically opposite the horseshoe and the outer ghost bubble
seen in X-ray maps, which could indicate that the inner jets once went
along a NW--SE axis rather than the current almost N--S one (for example,
the jets may precess \citealt{dunn2006}). There is a particularly
bright arc of X-ray emission in the SE direction between the Blue Loop and the nucleus, which means the X-ray
emitting gas is denser there. If this caused disruption of the jet
then the jet interaction may have been unusual and have led to more
cold gas being dragged out or the gas out there having been strongly
disturbed. We can find no particular reason why the Southern filament
shows star formation.

\section{Conclusions}

In this paper we present ACS observations of the extended star formation in NGC 1275, assumed to be associated with the low velocity system and tracing the spatial distribution of the H$\alpha$ filamentary system.

We find a bimodal distribution with average colours of $(\mathrm{F435W}-\mathrm{F625W})_{0}$ $\sim$ $-0.4$ to 0.1 and $(\mathrm{F435W}-\mathrm{F625W})_{0}$ $\sim$ $-0.1$ to 0.2 for the blue population in the Blue Loop and the Southern filament respectively. In the central region we find much redder colours, consistent with previous results. Assuming SSP models, these very blue colours translate into an age of less than 10$^{8}$ years, a factor of 10 younger than the youngest cluster populations in the core. This value is supported by assuming tidal interactions between these star forming regions and the centre of the galaxy.

Spatially the young blue populations fall on the bright arms of the H$\alpha$ loop and filament while the older, redder populations are more evenly dispersed around the galaxy. This and the extremely young ages implies the formation mechanism is both separate to the event that formed the star clusters in the centre of the galaxy and is intimately linked to the formation of the H$\alpha$ filaments.

The star formation in the Blue Loop and Southern filament is `clumpy' and there is evidence suggesting that there is a gradient in colours along these `clumps'. This, and the observation of a broad range in colours in the blue population suggest that the event responsible for the star cluster formation is unlikely to have occurred rapidly. However the range in ages of the population are very dependent on the models used and can only be precisely determined with spectroscopy.

We suggest formation mechanisms based on the disruption of the magnetic field supporting the filaments either as a shock from a buoyant radio bubble or by the filaments, fragmenting and falling back into the galaxy after being dragged out below such a bubble. If such a scenario is responsible for the star formation then these outer stellar regions provide another link in the evolutionary cycle feeding the SMBH and regulating the cluster growth. Spectroscopic observations of these regions are needed to confirm that the stellar clusters are really in the LVS and to better determine their ages and kinematics.

The star formation rate in the Blue Loop is $\sim20\mathrm{M}_{\odot}\,\mathrm{yr}^{-1}$. The lifetime of the whole low velocity filamentary system is therefore much greater than 10$^{8}\mathrm{yr}$.

\section{Acknowledgements}

REAC acknowledges STFC for financial support. ACF thanks the Royal Society. REAC would also like to thank Nate Bastian and Nina Hatch for helpful and interesting discussions and the anonymous referee for constructive criticism which has greatly improved this work.

This research has made use of the NASA/IPAC Extragalactic Database (NED) which is operated by the Jet Propulsion Laboratory, California Institute of Technology, under contract with the National Aeronautics and Space Administration.

\bibliographystyle{mnras}
\bibliography{mnras_template}

\begin{thebibliography}{}

\bibitem[\protect\citeauthoryear{{Ashman} \& {Zepf}}{{Ashman} \&
  {Zepf}}{1992}]{ashman1992}
{Ashman} K.~M.,  {Zepf} S.~E., 1992, \apj, 384, 50

\bibitem[\protect\citeauthoryear{{Boroson}}{{Boroson}}{1990}]{boroson1990}
{Boroson} T.~A., 1990, \apj, 360, 465

\bibitem[\protect\citeauthoryear{{Briggs}, {Snijders}, \&
  {Boksenberg}}{{Briggs} et~al.}{1982}]{briggs1982}
{Briggs} S.~A., {Snijders} M.~A.~J.,  {Boksenberg} A., 1982, \nat, 300, 336

\bibitem[\protect\citeauthoryear{{Brodie} et~al.}{{Brodie}
  et~al.}{1998}]{brodie1998}
{Brodie} J.~P., {Schroder} L.~L., {Huchra} J.~P., {Phillips} A.~C.,
  {Kissler-Patig} M.,  {Forbes} D.~A., 1998, \aj, 116, 691

\bibitem[\protect\citeauthoryear{{Brodie} \& {Strader}}{{Brodie} \&
  {Strader}}{2006}]{brodie2006}
{Brodie} J.~P.,  {Strader} J., 2006, \araa, 44, 193

\bibitem[\protect\citeauthoryear{{Bruzual} \& {Charlot}}{{Bruzual} \&
  {Charlot}}{2003}]{bc03}
{Bruzual} G.,  {Charlot} S., 2003, \mnras, 344, 1000

\bibitem[\protect\citeauthoryear{{Cardelli}, {Clayton}, \& {Mathis}}{{Cardelli}
  et~al.}{1989}]{cardelli1989}
{Cardelli} J.~A., {Clayton} G.~C.,  {Mathis} J.~S., 1989, \apj, 345, 245

\bibitem[\protect\citeauthoryear{{Carlson} et~al.}{{Carlson}
  et~al.}{1998}]{carlson1998}
{Carlson} M.~N. et~al., 1998, \aj, 115, 1778

\bibitem[\protect\citeauthoryear{{Caulet} et~al.}{{Caulet}
  et~al.}{1992}]{caulet1992}
{Caulet} A., {Woodgate} B.~E., {Brown} L.~W., {Gull} T.~R., {Hintzen} P.,
  {Lowenthal} J.~D., {Oliversen} R.~J.,  {Ziegler} M.~M., 1992, \apj, 388, 301

\bibitem[\protect\citeauthoryear{{Conselice}, {Gallagher}, \&
  {Wyse}}{{Conselice} et~al.}{2001}]{conselice2001}
{Conselice} C.~J., {Gallagher} J.~S., III,  {Wyse} R.~F.~G., 2001, \aj, 122,
  2281

\bibitem[\protect\citeauthoryear{{Crawford}}{{Crawford}}{2004}]{crawford2004}
{Crawford} C.~S., 2004, in {Mulchaey} J.~S., {Dressler} A.,  {Oemler} A., ed,
  Clusters of Galaxies: Probes of Cosmological Structure and Galaxy Evolution

\bibitem[\protect\citeauthoryear{{Dixon}, {Davidsen}, \& {Ferguson}}{{Dixon}
  et~al.}{1996}]{dixon1996}
{Dixon} W.~V.~D., {Davidsen} A.~F.,  {Ferguson} H.~C., 1996, \aj, 111, 130

\bibitem[\protect\citeauthoryear{{Donahue} et~al.}{{Donahue}
  et~al.}{2000}]{donahue2000}
{Donahue} M., {Mack} J., {Voit} G.~M., {Sparks} W., {Elston} R.,  {Maloney}
  P.~R., 2000, \apj, 545, 670

\bibitem[\protect\citeauthoryear{{Dunn}, {Fabian}, \& {Sanders}}{{Dunn}
  et~al.}{2006}]{dunn2006}
{Dunn} R.~J.~H., {Fabian} A.~C.,  {Sanders} J.~S., 2006, \mnras, 366, 758

\bibitem[\protect\citeauthoryear{{Ekers}, {van der Hulst}, \& {Miley}}{{Ekers}
  et~al.}{1976}]{ekers1976}
{Ekers} R.~D., {van der Hulst} J.~M.,  {Miley} G.~K., 1976, \nat, 262, 369

\bibitem[\protect\citeauthoryear{{Fabian} et~al.}{{Fabian}
  et~al.}{2008}]{fabian2008}
{Fabian} A.~C., {Johnstone} R.~M., {Sanders} J.~S., {Conselice} C.~J.,
  {Crawford} C.~S., {Gallagher} J.~S., III,  {Zweibel} E., 2008, \nat, 454, 968

\bibitem[\protect\citeauthoryear{{Fabian} \& {Nulsen}}{{Fabian} \&
  {Nulsen}}{1977}]{fabian1977}
{Fabian} A.~C.,  {Nulsen} P.~E.~J., 1977, \mnras, 180, 479

\bibitem[\protect\citeauthoryear{{Fabian} et~al.}{{Fabian}
  et~al.}{2003}]{fabian2003}
{Fabian} A.~C., {Sanders} J.~S., {Crawford} C.~S., {Conselice} C.~J.,
  {Gallagher} J.~S.,  {Wyse} R.~F.~G., 2003, \mnras, 344, L48

\bibitem[\protect\citeauthoryear{{Fabian} et~al.}{{Fabian}
  et~al.}{2000}]{fabian2000}
{Fabian} A.~C. et~al., 2000, \mnras, 318, L65

\bibitem[\protect\citeauthoryear{{Fabian} et~al.}{{Fabian}
  et~al.}{2006}]{fabian2006}
{Fabian} A.~C., {Sanders} J.~S., {Taylor} G.~B., {Allen} S.~W., {Crawford}
  C.~S., {Johnstone} R.~M.,  {Iwasawa} K., 2006, \mnras, 366, 417

\bibitem[\protect\citeauthoryear{{Ferruit} et~al.}{{Ferruit}
  et~al.}{1997}]{ferruit1997}
{Ferruit} P., {Adam} G., {Binette} L.,  {Pecontal} E., 1997, New Astronomy, 2,
  345

\bibitem[\protect\citeauthoryear{{Ferruit} \& {Pecontal}}{{Ferruit} \&
  {Pecontal}}{1994}]{ferruit1994}
{Ferruit} P.,  {Pecontal} E., 1994, \aap, 288, 65

\bibitem[\protect\citeauthoryear{{Forbes}, {Brodie}, \& {Grillmair}}{{Forbes}
  et~al.}{1997}]{forbes1997}
{Forbes} D.~A., {Brodie} J.~P.,  {Grillmair} C.~J., 1997, \aj, 113, 1652

\bibitem[\protect\citeauthoryear{{Gillmon}, {Sanders}, \& {Fabian}}{{Gillmon}
  et~al.}{2004}]{gillmon2004}
{Gillmon} K., {Sanders} J.~S.,  {Fabian} A.~C., 2004, \mnras, 348, 159

\bibitem[\protect\citeauthoryear{{Hatch} et~al.}{{Hatch}
  et~al.}{2005}]{hatch2005}
{Hatch} N.~A., {Crawford} C.~S., {Fabian} A.~C.,  {Johnstone} R.~M., 2005,
  \mnras, 358, 765

\bibitem[\protect\citeauthoryear{{Hatch} et~al.}{{Hatch}
  et~al.}{2006}]{hatch2006}
{Hatch} N.~A., {Crawford} C.~S., {Johnstone} R.~M.,  {Fabian} A.~C., 2006,
  \mnras, 367, 433

\bibitem[\protect\citeauthoryear{{Heckman} et~al.}{{Heckman}
  et~al.}{1989}]{heckman1989}
{Heckman} T.~M., {Baum} S.~A., {van Breugel} W.~J.~M.,  {McCarthy} P., 1989,
  \apj, 338, 48

\bibitem[\protect\citeauthoryear{{Ho}, {Lim}, \& {Dinh-V-Trung}}{{Ho}
  et~al.}{2009}]{ho2009}
{Ho} I.-T., {Lim} J.,  {Dinh-V-Trung} , 2009, \apj, 698, 1191

\bibitem[\protect\citeauthoryear{{Holtzman} et~al.}{{Holtzman}
  et~al.}{1992}]{holtzman1992}
{Holtzman} J.~A. et~al., 1992, \aj, 103, 691

\bibitem[\protect\citeauthoryear{{Holtzman} et~al.}{{Holtzman}
  et~al.}{1995}]{holtzman1995}
{Holtzman} J.~A. et~al., 1995, \pasp, 107, 156

\bibitem[\protect\citeauthoryear{{Hu} et~al.}{{Hu} et~al.}{1983}]{hu1983}
{Hu} E.~M., {Cowie} L.~L., {Kaaret} P., {Jenkins} E.~B., {York} D.~G.,
  {Roesler} F.~L., 1983, \apjl, 275, L27

\bibitem[\protect\citeauthoryear{{Inoue} et~al.}{{Inoue}
  et~al.}{1996}]{inoue1996}
{Inoue} M.~Y., {Kameno} S., {Kawabe} R., {Inoue} M., {Hasegawa} T.,  {Tanaka}
  M., 1996, \aj, 111, 1852

\bibitem[\protect\citeauthoryear{{Johnstone} \& {Fabian}}{{Johnstone} \&
  {Fabian}}{1988}]{johnstone1988}
{Johnstone} R.~M.,  {Fabian} A.~C., 1988, \mnras, 233, 581

\bibitem[\protect\citeauthoryear{{Keel} \& {White}}{{Keel} \&
  {White}}{2001}]{keel2001}
{Keel} W.~C.,  {White} R.~E., III, 2001, \aj, 121, 1442

\bibitem[\protect\citeauthoryear{{Kent} \& {Sargent}}{{Kent} \&
  {Sargent}}{1979}]{kent1979}
{Kent} S.~M.,  {Sargent} W.~L.~W., 1979, \apj, 230, 667

\bibitem[\protect\citeauthoryear{{Kotulla} et~al.}{{Kotulla}
  et~al.}{2009}]{Kotulla2009}
{Kotulla} R., {Fritze} U., {Weilbacher} P.,  {Anders} P., 2009, \mnras, 396,
  462

\bibitem[\protect\citeauthoryear{{Lazareff} et~al.}{{Lazareff}
  et~al.}{1989}]{lazareff1989}
{Lazareff} B., {Castets} A., {Kim} D.-W.,  {Jura} M., 1989, \apjl, 336, L13

\bibitem[\protect\citeauthoryear{{Lim}, {Ao}, \& {Dinh-V-Trung}}{{Lim}
  et~al.}{2008}]{lim2008}
{Lim} J., {Ao} Y.,  {Dinh-V-Trung} , 2008, \apj, 672, 252

\bibitem[\protect\citeauthoryear{{Lynds}}{{Lynds}}{1970}]{lynds1970}
{Lynds} R., 1970, \apjl, 159, L151

\bibitem[\protect\citeauthoryear{{McKee} et~al.}{{McKee}
  et~al.}{1993}]{mckee1993}
{McKee} C.~F., {Zweibel} E.~G., {Goodman} A.~A.,  {Heiles} C., 1993, in {Levy}
  E.~H.,  {Lunine} J.~I., ed, Protostars and Planets III, p. 327

\bibitem[\protect\citeauthoryear{{McNamara} \& {Nulsen}}{{McNamara} \&
  {Nulsen}}{2007}]{mcnamara2007}
{McNamara} B.~R.,  {Nulsen} P.~E.~J., 2007, \araa, 45, 117

\bibitem[\protect\citeauthoryear{{McNamara}, {O'Connell}, \&
  {Sarazin}}{{McNamara} et~al.}{1996}]{mcnamara1996}
{McNamara} B.~R., {O'Connell} R.~W.,  {Sarazin} C.~L., 1996, \aj, 112, 91

\bibitem[\protect\citeauthoryear{{Minkowski}}{{Minkowski}}{1955}]{minkowski195%
5}
{Minkowski} R., 1955, Carnegie Yearbook, 54, 25

\bibitem[\protect\citeauthoryear{{Minkowski}}{{Minkowski}}{1957}]{minkowski195%
7}
{Minkowski} R., 1957, in IAU Symposium, Vol.~4, {van de Hulst} H.~C., ed, Radio
  astronomy, p. 107

\bibitem[\protect\citeauthoryear{{Norgaard-Nielsen} et~al.}{{Norgaard-Nielsen}
  et~al.}{1993}]{norgaardnielsen1993}
{Norgaard-Nielsen} H.~U., {Goudfrooij} P., {Jorgensen} H.~E.,  {Hansen} L.,
  1993, \aap, 279, 61

\bibitem[\protect\citeauthoryear{{Pei}}{{Pei}}{1992}]{pei1992}
{Pei} Y.~C., 1992, \apj, 395, 130

\bibitem[\protect\citeauthoryear{{Peterson} \& {Fabian}}{{Peterson} \&
  {Fabian}}{2006}]{peterson2006}
{Peterson} J.~R.,  {Fabian} A.~C., 2006, Phys. Rep., 427, 1

\bibitem[\protect\citeauthoryear{{Richer} et~al.}{{Richer}
  et~al.}{1993}]{richer1993}
{Richer} H.~B., {Crabtree} D.~R., {Fabian} A.~C.,  {Lin} D.~N.~C., 1993, \aj,
  105, 877

\bibitem[\protect\citeauthoryear{{Salom{\'e}} et~al.}{{Salom{\'e}}
  et~al.}{2006}]{salome2006}
{Salom{\'e}} P. et~al., 2006, \aap, 454, 437

\bibitem[\protect\citeauthoryear{{Salom{\'e}} et~al.}{{Salom{\'e}}
  et~al.}{2008a}]{salome2008}
{Salom{\'e}} P., {Combes} F., {Revaz} Y., {Edge} A.~C., {Hatch} N.~A., {Fabian}
  A.~C.,  {Johnstone} R.~M., 2008a, \aap, 484, 317

\bibitem[\protect\citeauthoryear{{Salom{\'e}} et~al.}{{Salom{\'e}}
  et~al.}{2008b}]{salome2008a}
{Salom{\'e}} P., {Revaz} Y., {Combes} F., {Pety} J., {Downes} D., {Edge} A.~C.,
   {Fabian} A.~C., 2008b, \aap, 483, 793

\bibitem[\protect\citeauthoryear{{Sandage}}{{Sandage}}{1971}]{sandage1971}
{Sandage} A.~R., 1971, in {O'Connell} D.~J.~K., ed, Study Week on Nuclei of
  Galaxies, p. 271

\bibitem[\protect\citeauthoryear{{Sanders} \& {Fabian}}{{Sanders} \&
  {Fabian}}{2007}]{sanders2007}
{Sanders} J.~S.,  {Fabian} A.~C., 2007, \mnras, 381, 1381

\bibitem[\protect\citeauthoryear{{Sanders} et~al.}{{Sanders}
  et~al.}{2004}]{sanders2004}
{Sanders} J.~S., {Fabian} A.~C., {Allen} S.~W.,  {Schmidt} R.~W., 2004, \mnras,
  349, 952

\bibitem[\protect\citeauthoryear{{Schlegel}, {Finkbeiner}, \&
  {Davis}}{{Schlegel} et~al.}{1998}]{schlegel1998}
{Schlegel} D.~J., {Finkbeiner} D.~P.,  {Davis} M., 1998, \apj, 500, 525

\bibitem[\protect\citeauthoryear{{Shields}, {Filippenko}, \& {Basri}}{{Shields}
  et~al.}{1990}]{shields1990}
{Shields} J.~C., {Filippenko} A.~V.,  {Basri} G., 1990, \aj, 100, 1805

\bibitem[\protect\citeauthoryear{{Sirianni} et~al.}{{Sirianni}
  et~al.}{2005}]{sirianni2005}
{Sirianni} M. et~al., 2005, \pasp, 117, 1049

\bibitem[\protect\citeauthoryear{{Tomisaka}, {Ikeuchi}, \&
  {Nakamura}}{{Tomisaka} et~al.}{1988}]{tomisaka1988}
{Tomisaka} K., {Ikeuchi} S.,  {Nakamura} T., 1988, \apj, 335, 239

\bibitem[\protect\citeauthoryear{{Trancho} et~al.}{{Trancho}
  et~al.}{2006}]{trancho2006}
{Trancho} G., {Miller} B., {Garc{\'{\i}}a-Lorenzo} B.,  {S{\'a}nchez} S.~F.,
  2006, New Astronomy Review, 49, 613

\bibitem[\protect\citeauthoryear{{Unger} et~al.}{{Unger}
  et~al.}{1990}]{unger1990}
{Unger} S.~W., {Taylor} K., {Pedlar} A., {Ghataure} H.~S., {Penston} M.~V.,
  {Robinson} A., 1990, \mnras, 242, 33P

\end{thebibliography}

\end{document}